\documentclass[11pt,a4paper]{article}
\usepackage{amssymb,amsfonts,amsmath}
\usepackage{upgreek}
\usepackage[cm]{fullpage}
\usepackage{multicol}
\usepackage[labelfont=it,hypcap=true,labelformat=simple]{subcaption}
\newcommand{\sublabel}[1]{\caption{}\label{#1}}
\usepackage{graphicx}
\usepackage{mdframed}

\def\endarticle{\endtwocolumns\newpage
\expandafter\gdef\csname table*\endcsname{\@dblfloat{table}}
\expandafter\gdef\csname endtable*\endcsname{\end@dblfloat}
\expandafter\gdef\csname figure*\endcsname{\@dblfloat{figure}}
\expandafter\gdef\csname endfigure*\endcsname{\end@dblfloat}
\gdef\figure{\futurelet\next\lookforbrac}}

\newlength{\figheight}
\newlength{\figwidth}
\newlength{\margincrunch}
\newcommand{\fig}[2][]{Fig.\ \ref{#2}#1}
\newcommand{\sfig}[2][]{Fig.\ \ref{#2}#1}

\newcommand{\eq}[2][]{Eq.\ \textbf{\ref{#2}#1}}

\newcommand{\SIname}{Appendix}
\newcommand{\refsup}[1]{\SIname\ \ref{#1}} 
\newcommand{\citetext}[1]{\cite{#1}}

\usepackage{amsmath,amssymb}
\usepackage{color}

\newcommand{\trsp}{^{\mathrm{T}}}

\newcommand{\vecteur}[1]{{#1}}

\newcommand{\pt}[1]{\boldsymbol{#1}}
\newcommand{\vc}[1]{\pt{\vecteur{#1}}}

\newcommand{\ts}[1]{\boldsymbol{\mathbf{#1}}}

\newcommand{\grad}{\vc{\nabla}}

\renewcommand{\div}{\grad\cdot}

\newcommand{\ded}[2]{\frac{\partial #1}{\partial #2}}

\newcommand{\dedt}[1]{\frac{\partial #1}{\partial t}}

\usepackage{upgreek}

\newcommand{\micron}{\mathrm{\upmu m}}

\renewcommand{\ts}[1]{\boldsymbol{#1}}

\newcommand{\lded}[2]{{\partial #1}/{\partial #2}}

\newcommand{\ucm}[1]{\stackrel{\triangledown}{#1}}

\newcommand{\at}[1]{\!\left|_{#1}\right.}

\newcommand{\es}{\vc{e}_s}
\newcommand{\ep}{\vc{e}_{\phi}}
\newcommand{\en}{\vc{e}_n}

\newcommand{\tsigma}{\ts{\sigma}}

\newcommand{\eps}{{{\varepsilon}}}
\newcommand{\deps}{\dot{\eps}}

\newcommand{\teps}{\ts{{\eps}}}

\newcommand{\Rc}{R_c}

\newcommand{\Ta}{\tau_{\alpha}}
\newcommand{\Tc}{\tau_{\rm c}}
\newcommand{\tSigma}{\ts{\Sigma}}
\newcommand{\vp}{v_{\mathrm{t}}}
\newcommand{\va}{v_{\alpha}}

\newcommand{\sig}{\sigma}

\newcommand{\sigO}{\sig_{\perp}}

\newcommand{\zet}{\zeta}

\newcommand{\Sig}{\Sigma}
\newcommand{\SigO}{\Sigma_{\perp}}

\newcommand{\lfrac}[2]{#1/(#2)}
\newcommand{\ARE}{\vc{R}}

\begin{document}
\def\usebibtex{1}


\newcommand{\affilset}[2]{%
   \footnote{#2}
    \newcounter{#1}
    \setcounter{#1}{\value{footnote}}%
}
\newcommand{\affil}[1]{%
    \footnotemark[\value{#1}]%
}
\def\andname{,}
\title{Collective dynamics of actomyosin cortex\\endow cells with intrinsic mechanosensing properties}
\author{Jocelyn \'Etienne,\affilset{1}{Univ.\ Grenoble Alpes, LIPHY, F-38000 Grenoble, France}\affilset{2}{CNRS, LIPHY, F-38000 Grenoble, France}\affilset{4}{to whom correspondence should be addressed; E-mail: Jocelyn.Etienne@ujf-grenoble.fr}
Jonathan Fouchard, 
D\'emosth\`ene Mitrossilis,
Nathalie Bufi,\and
Pauline Durand-Smet, and
Atef Asnacios\affilset{3}{Universit\'e Paris-Diderot--CNRS, Laboratoire Mati\`ere et Syst\`emes Complexes (MSC), UMR 7057, Paris, France}}

\date{Accepted for publication in the \\\emph{Proceedings of the National Academy of Sciences of the USA} \\in a revised version}






\maketitle


\medskip

\begin{abstract} 
Living cells adapt and respond actively to the mechanical properties of their
environment.  In addition to biochemical mechanotransduction, evidence exists for a
myosin-dependent, purely mechanical sensitivity to the stiffness of the surroundings
at the scale of the whole cell.
Using a minimal model of the dynamics of actomyosin cortex,
we show that the interplay of myosin power strokes with the rapidly 
remodelling actin network
results in a regulation of force and cell shape that adapts to the
stiffness of the environment. Instantaneous changes of the environment
stiffness are found to trigger an intrinsic mechanical response of the actomyosin cortex.
Cortical retrograde flow resulting from actin polymerisation at the edges is shown to
be modulated by the stress resulting from myosin contractility, which in turn
regulates the cell size in a force-dependent manner. 
%
The model describes the maximum force that cells can exert and the maximum speed at
which they can contract, which are measured experimentally.  These limiting cases are found to
be associated with energy dissipation phenomena which are of the same nature as those
taking place during the contraction of a whole muscle.  This explains the fact that
single nonmuscle cell and whole muscle contraction both follow a Hill-like
force--velocity relationship.  
\end{abstract}

\smallskip
\begin{multicols}{2}
{W}hen placed in different mechanical environments, living cells assume
different shapes
\cite{Turner-Mahowald.1977.1,Engler+Discher.2004.1,Yeung+Janmey.2005.1,Saez+Ladoux.2007.1,Zhong+Fabry-ONeill.2012.1}.
This behaviour is known to be strongly dependent on the contractile activity of
the acto\-myosin network
\cite{Young+Kiehart.1993.1,Pelham-Wang.1997.1,Chicurel+Ingber.1998.1,Zajac-Discher.2008.1,Cai+Sheetz.2010.1}.
One of the cues driving the cell response to its environment is rigidity
\cite{Discher+.2005.1}.  Cells are able to sense not only
the local rigidity of the material they are in contact with \cite{Vogel-Sheetz.2006.1}, but also the one
associated with distant cell-substrate contacts :
that is, the amount of extra force needed in order to achieve a given
displacement of microplates between which the cell is placed
\cite{Mitrossilis+Asnacios.2009.1}, \fig{fig.atef.manip}, 
of an AFM cantilever \cite{Webster+Fletcher.2011.1,Crow+Fletcher.2012.1},
or of elastic micropillars \cite{Trichet+Voituriez-Ladoux.2012.1}. 
This cell-scale 
rigidity sensing is totally dependent on Myosin-II activity \cite{Mitrossilis+Asnacios.2009.1}.
A working model of the molecular mechanisms at play in the actomyosin cortex is
available \cite{Rossier+Sheetz.2010.1}, where myosin contraction, actin
treadmilling and actin crosslinker turn\-over are the main ingredients.
Phenomenological models
\cite{Zemel+Discher-Safran.2010.1,Marcq+Prost.2011.1} of mechanosensing have
been proposed, but could not bridge the gap between the molecular microstructure and this
phenomenology. Here, we show that the collective dynamics of
actin, actin crosslinkers and myosin molecular motors are sufficient to explain 
cell-scale rigidity sensing. 
The model derivation is analogous to the one of rubber elasticity of transiently crosslinked networks
\cite{Yamamoto.1956.1}, with the addition of active crosslinkers, accounting for the myosin.
It involves four parameters only: myosin contractile stress,
speed of actin treadmilling, elastic modulus, and viscoelastic relaxation time of the cortex, which 
arises because of crosslinker unbinding. 
It allows quantitative predictions of
the dynamics and statics of cell contraction depending on the external stiffness.
The crucial dependence of this behaviour on the fact that
crosslinkers have a short life time is reminiscent of the model
of muscle contraction by A. F. Huxley \cite{Huxley.1957.1}, in 
which the force dependence of muscle contraction rate is explained by
the fact that for lower muscle force and higher contraction speed, the
number of myosin heads contributing to filament sliding decreases in
favour of those resisting it transiently, before they unbind.
While it is known that many molecules associated with actomyosin exhibit
stress-dependent dynamics \cite{Kovacs+Sellers.2007.1}, collective effects alone
can explain observations in both Huxley's muscle contraction model and the present model.
We have previously evidenced the similarity of single nonmuscle cell and muscle
contraction rate \cite{Mitrossilis+Asnacios.2009.1}. In spite of very dissimilar organisation
of actomyosin in muscles, where it forms well-ordered sarcomeres, and in nonmuscle cell cortex,
where no large-scale patterning is observed, we show that corresponding mechanisms explain their similar motor
properties.
The collective dynamics we describe are consistent with the fact that the actin network behaves as a fluid
at long times.
We show how myosin activity can contract this fluid at a given rate
that depends on external forces resisting cell contraction, arising e.g.\  
from the stiffness of the environment. This, combined with actin
protrusivity, results in both a sustained retrograde flow and a
regulation of cell shape.
In addition, this explains the elastic-like
behaviour observed in cell-scale rigidity-sensing and justifies \emph{ad hoc}
models based on this observation \cite{Zemel+Discher-Safran.2010.1}.

\newlength{\figtotwidth}
\setlength{\figtotwidth}{.9\textwidth}
\setlength{\figwidth}{.28\figtotwidth}
\setlength{\figheight}{.28\figtotwidth}
\setlength{\margincrunch}{.0\figwidth}
\addtolength{\margincrunch}{-.0\linewidth}
\begin{figure*}[tbp]
\makeatletter
\hspace*{-.55\margincrunch}
\begin{tabular}{lll}
\begin{subfigure}[t]{1.4\figwidth}
 \centering
 \begin{minipage}[b]{1.4\figwidth}
 \includegraphics[height=\figheight,origin=tl]{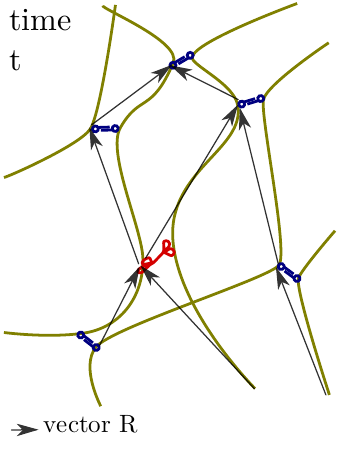}%
 \includegraphics[height=\figheight,origin=tl]{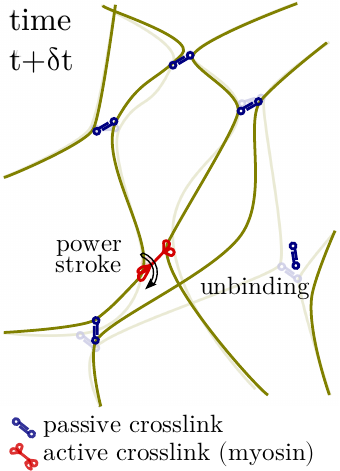}
 \end{minipage}
\sublabel{fig.atef.network}
\end{subfigure}
&
\begin{subfigure}[t]{1.15\figwidth}
 \begin{minipage}[b]{1.15\figwidth}
 \centering\includegraphics[height=1.15\figheight,origin=tl]{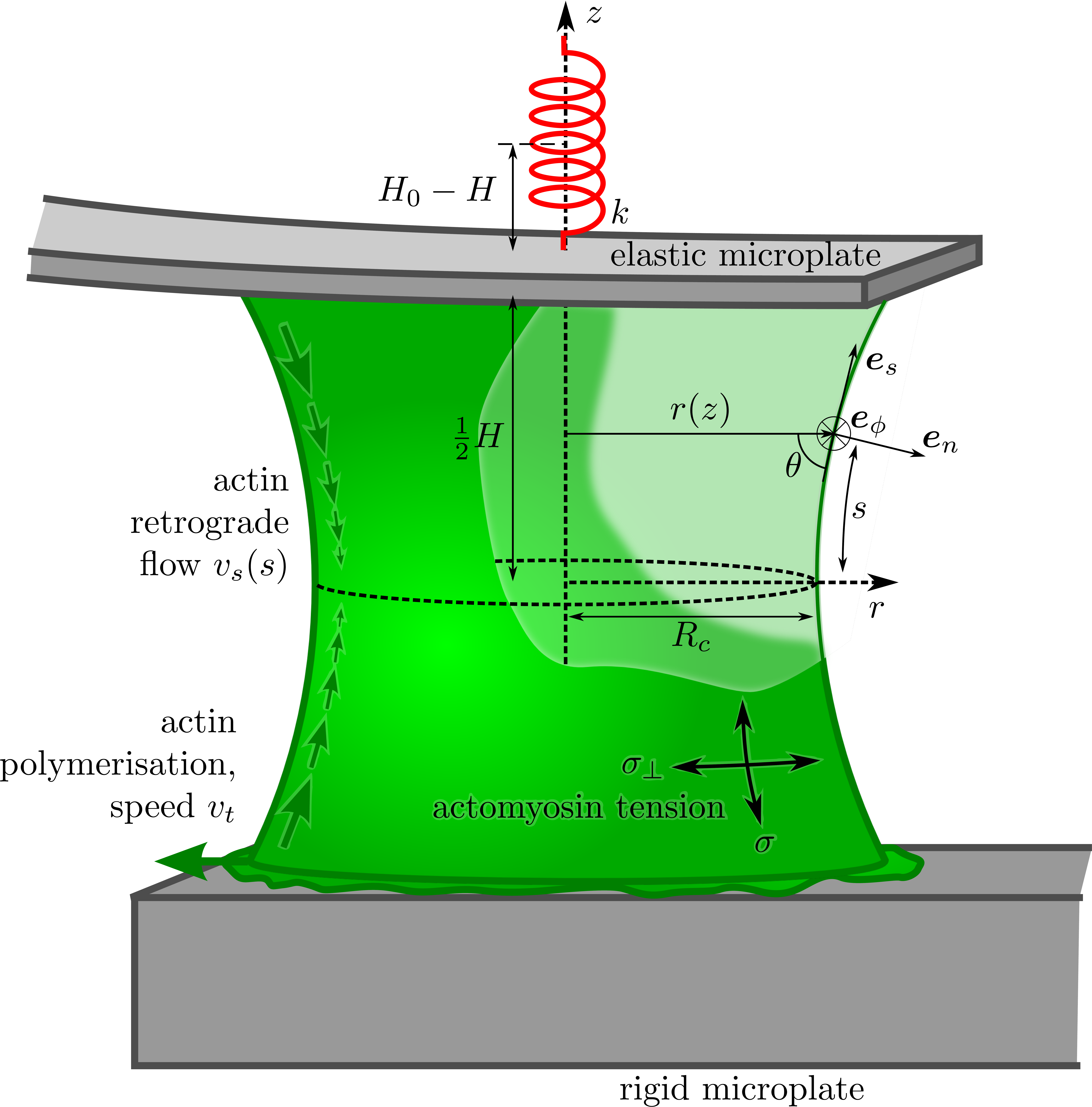}
 \end{minipage}
\sublabel{fig.atef.sketch}
\end{subfigure}
&
\begin{subfigure}[t]{\figwidth}
 \centering\includegraphics[height=\figheight]{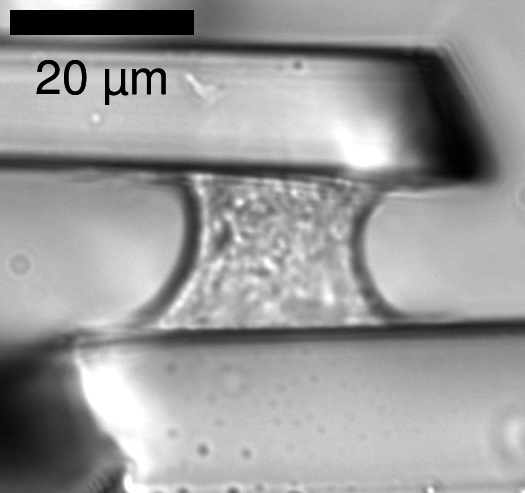}
\sublabel{fig.atef.manip}
\end{subfigure}
\end{tabular}
\hspace*{-.55\margincrunch}
\caption{Model of the acto\discretionary {-}{}{}myosin behaviour and experimental setup. (\emph{\subref {fig.atef.network}}) Sketch of a transiently crosslinked actin network with myosin bipolar filaments. Unbinding of a crosslinker releases elastic tension locally, the crosslinker will re-bind to the network, preserving its elastic properties, but after this loss of stored elastic energy. Myosin power-strokes have the effect of modifying the equilibrium length of the adjacent strands $\boldsymbol {{R}}$, this results in an increase of the tension (see Supplementary Discussion \ref {sup.model}). (\emph{\subref {fig.atef.sketch}})~Model of the mechanical components of the cell and microplate system. Microplates impose that the vertical force equilibrating the cortical tension is linked to the cell half-height $h$ with $F=2k(h_0-h)$. Tension along the acto\discretionary {-}{}{}myosin cortex (green surface) is anisotropic and has values $\sigma $ and $\sigma _{\perp }$ along directions $\boldsymbol {{e}}_s$ and $\boldsymbol {{e}}_{\phi }$. Actin treadmilling provides a boundary condition at the cell leading edge, the actin cortex undergoes a retrograde flow away from the plates. (\emph{\subref {fig.atef.manip}})~Transmission image of a cell and setup.}
\label{fig.atef}
\end{figure*}

\section*{Results}
\subsection*{Intrinsic rigidity sensing of actomyosin}
The actin cortex of nonmuscle cells is a disordered network situated at the cell periphery. 
Actin filaments are crosslinked by proteins such as $\alpha$-actinin, however these crosslinkers experience
a rapid turnover, of order $10$~s e.g.\ for $\alpha$-actinin
\cite{Mukhina+.2007.1}, and actin itself has a scarcely longer turnover time
\cite{Fritzsche+Charras.2013.1}. The actin network is thus only transiently crosslinked.
Following \cite{Yamamoto.1956.1}, we describe the behaviour of such a network by a rubber-like model. 
Up to the first order, this model leads to a stress-strain relationship of a Maxwell viscoelastic liquid in which
the relaxation time is a characteristic unbinding time $\Ta$, 
\begin{align}
\Ta \ucm{\tsigma} + \tsigma - 2 \Ta E \dot{\teps} = \vc{0},
\label{eq.cst_nomyo}
\end{align}
where $\dot{\teps}$ is the rate-of-strain tensor and
$\ucm{\tsigma}$ the objective time-derivative of the stress tensor $\tsigma$, see \refsup{sup.model}. 
In the linear setting, $\tsigma=2E(\beta^2\left<\ARE\ARE\right>-\ts{I})$,
where $E$ is the elastic modulus of the crosslinked actin network,
$\beta$ is inversely proportional to the Kuhn length, and $\ARE$ is
a basic element of this network, namely the strand vector spanning the distance between any two consecutive
actin--actin bonds, \fig{fig.atef.network}. 
As long as these two bonds hold (for times much shorter than $\Ta$), 
this basic element behaves elastically, and 
the stress tensor $\tsigma$ grows in proportion with the strain.
When a crosslinker unbinds, the filaments can slide
somewhat relative to one another, and the elastic tension that was maintained
{via} this crosslinker is relaxed: this corresponds to an effective friction,
and happens at a typical rate $1/\Ta$.
In sum,
the actin network behaves like an elastic solid of modulus $E$ over a time shorter than $\Ta$,
as crosslinkers remain in place during such a solicitation,
and it has a viscous-like response for longer times with an effective viscosity $\Ta E$, 
since the network yields as crosslinkers unbind. 

Such a viscoelastic liquid is unable to resist mechanical 
stresses \cite{Vaziri-Gopinath.2008.1}. However living cells are able to generate stress themselves
\cite{Harris+.1981.1} thanks to myosin bipolar filaments, which act as actin crosslinkers
but are in addition able to move by one step length 
along one of the filaments they are bound to, using biochemical energy.
Let us call $\alpha_{\text{myo}}$ the fraction of crosslinkers which are myosin filaments, 
and effectuate a power-stroke of step length $\ell$ at frequency $1/\tau_{\text{myo}}$. 
The power-stroke corresponds to a
change of the binding location of the myosin head along the actin filament,
and thus
affects the length of the neighbouring strands $\ARE$,  \fig{fig.atef.network}. Supplementary \eq{eq.proba} includes
the additional term that describes this myosin-driven evolution of the network 
configuration. When this equation is integrated to give the local macroscopic stress tensor
$\tsigma$, this term results in 
a contractile stress $\tSigma$ proportional to ${\Ta}E\alpha_{\text{myo}}(\ell\beta)^2/{\tau_{\text{myo}}} $
(see \refsup{sup.model}):
\begin{align}
\Ta \ucm{\tsigma} + \tsigma - 2 \Ta E \dot{\teps} = \tSigma.
\label{eq.cst} 
\end{align} 
The three-parameter model obtained ($\Ta$, $E$, $\tSigma$) is in line with early continuum models
\cite{He-Dembo.1997.1} and the active gel theory \cite{Kruse+.2005.1}, however
we do not supplement this active stress with an elastic stress,
unlike previous models of mechanosensitive active gels
\cite{Marcq+Prost.2011.1,Trichet+Voituriez-Ladoux.2012.1} where cells
are treated as viscoelastic \emph{solids} (\refsup{sup.othermodels}).
The interpretation of this equation is that the contractile
stress $\tSigma$ gives rise either to the build-up of a contractile
tension $\tsigma$ (if clamped boundary conditions allow no strain) or a contractile strain rate
$\tSigma/(2\Ta E)$
(if free boundary conditions allow strain but not tension build-up, this is e.g.\ the case 
of super-precipitation \emph{in vitro} \cite{Soares+Koenderink.2012.1}), or a combination of these.

\setlength{\figwidth}{.3845\figtotwidth}
\setlength{\figheight}{.71\figwidth}
\addtolength{\margincrunch}{-\linewidth}
\begin{figure*}[tbp]
\begin{tabular}{lll}
\begin{subfigure}[b]{\figwidth}
 \includegraphics[height=\figheight,origin=tl]{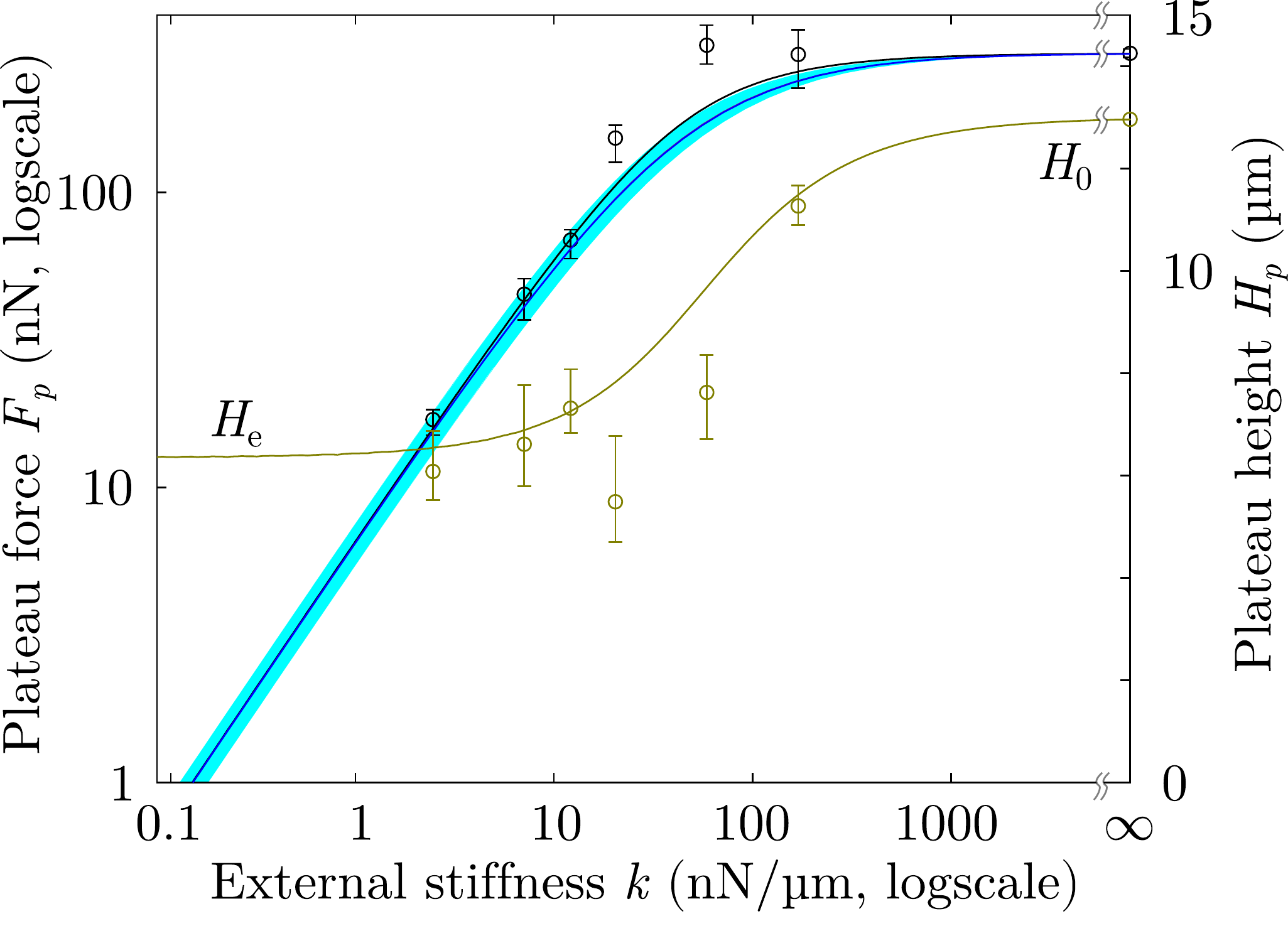}
\sublabel{fig.F_vs_k}
\end{subfigure}
&
\begin{subfigure}[b]{.9\figwidth}
 \includegraphics[height=\figheight,origin=tl]{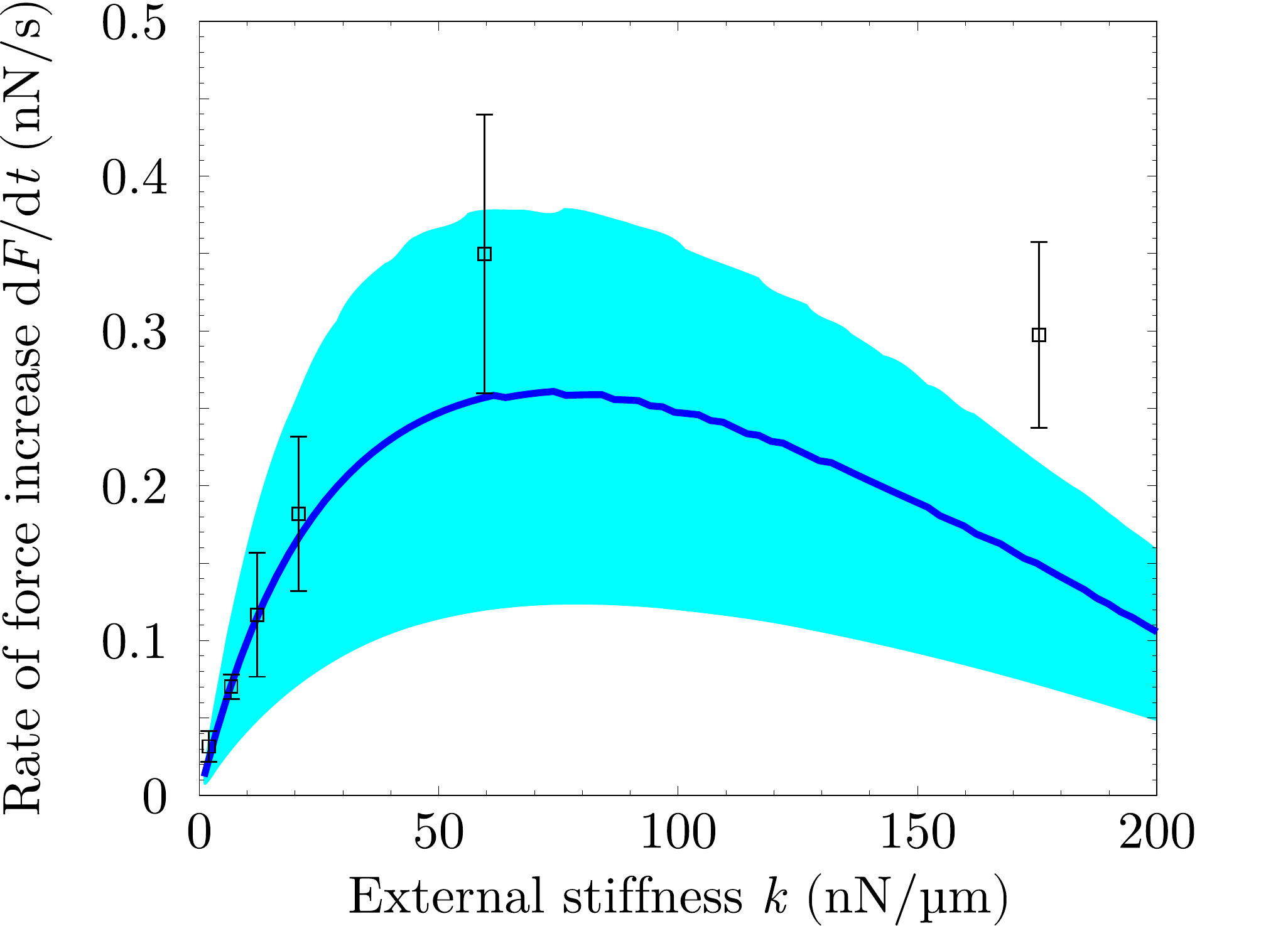}
\sublabel{fig.dFdt_vs_k}
\end{subfigure}
&
\begin{subfigure}[b]{.7\figwidth}
 \centering
 \includegraphics[height=\figheight,origin=tl]{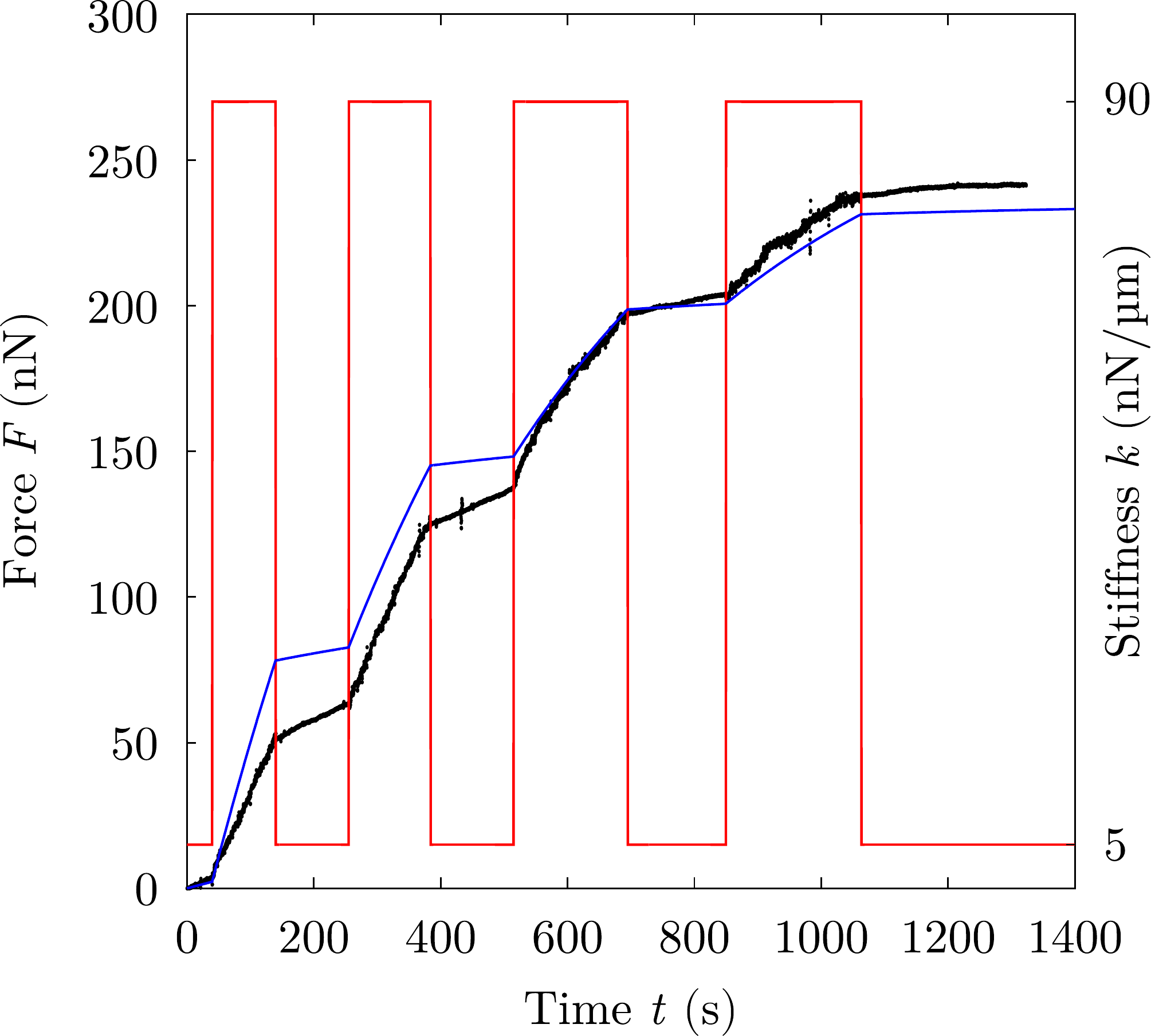}
\sublabel{fig.change_k}
\end{subfigure}
\end{tabular}
\caption{Predictive modelling of the stiffness-dependent cell mechanical response. (\emph{\subref {fig.F_vs_k}}) The force and cell height at equilibrium are adapted to the stiffness of the environment up to a maximum force at high stiffness, and, for vanishing stiffness, there is a well defined equilibrium length $H_e$ (independent of microtubules, see \sfig{fig.microtubules}). Circles, experimental results \cite {Mitrossilis+Asnacios.2009.1} for force (black) and height (green), black line, force predicted by the 3D model, green line, height predicted by the 3D model, blue curve and shaded area, force and confidence interval in the 1D model. Two out of the four parameters of the model adjusted in this plot, using the force at infinite stiffness and the height at zero stiffness. (\emph{\subref {fig.dFdt_vs_k}}) During the transient part of the experiment, the rate of loading of the cell is adapted to the stiffness of the environment. Boxes, experimental results \cite {Mitrossilis+Asnacios.2009.1}, green curve and shaded area, force and confidence interval in the 1D model. One parameter of the model is fitted in this plot, the last one is adjusted on figure \ref {fig.hill_values}. (\emph{\subref {fig.change_k}}) Instantaneous adaptation to a change of the microplate stiffness $k$. Red line, stiffness imposed using a feedback loop, black dots, force measured \cite {Mitrossilis+Asnacios.2010.1}, blue line, 1D model prediction of force, using the stiffness changes imposed in the experiment (red line), and the four parameters obtained in ({\subref {fig.F_vs_k}}) and ({\subref {fig.dFdt_vs_k}}), without any further adjustment.}
\label{fig.data}
\end{figure*}

We then asked whether this simple rheological law for the actomyosin cortex
could explain the behaviour of cells in our microplate experiments, \fig{fig.atef.manip}. To do so, we
investigated the equilibrium shape and force of a thin actomyosin cortex surrounding
a liquid (the cytosol), in the three-dimensional geometry of the experimental setup
in \fig{fig.atef.sketch}, \refsup{sup.three_d}. Surprisingly, the rigidity sensing
property of cells is adequately recovered by this simple model: while a fixed
maximum force is predicted above a certain critical stiffness $k_c=\Sigma S/H_0$ of the
microplates, the actomyosin-generated force is proportional to $k<k_c$, \fig{fig.F_vs_k}. 
Here $H_0$ is the initial plate separation and $S$ is the section area of the actin network. 
Thus the contractile activity of myosin motors is enough to endow the
viscoelastic liquid-like actin cortex with the spring-like response to the
rigidity of its environment \cite{Saez+Ladoux.2005.1,Mitrossilis+Asnacios.2009.1}, 
a property which was introduced phenomenologically in previous models 
\cite{Zemel+Discher-Safran.2010.1}.
In order to get a clear understanding of the mechanism through which this was
possible, we simplified the geometry to a one-dimensional problem
(\fig{fig.rheo.actin}) and found that the spring-like behaviour of the
contractile fluid was retained, \fig{fig.rheo.plot} and \refsup{sup.one_d}.
Indeed, for an environment (external spring) of stiffness $k$ beyond the
critical value $k_c$, the
contractile fluid is unable to strain beyond some plate deflection smaller than $H_0$,
and equilibrium is reached as it exerts its maximal contractility
$\Sigma$. For an external stiffness $k$ below the critical value, the tension $\sigma$ 
which is balanced by microplate force is smaller than $\Sigma$ for any admissible 
deflection. Thus there is a nonzero rate of contraction, $\deps<0$, as long as a
maximal deflection is not reached, which is $H_0$ with the hypotheses done so far. 
The next section discusses actin polymerisation as limiting this deflection,
thus setting an equilibrium cell size.

The force finally achieved is proportional to $k$ --- just as a prestretched spring of stiffness
$k_c$ would do (\fig{fig.rheo.spring}), or, alternatively, just as cells do
when exhibiting a mechanosensitive behaviour
\cite{Zemel+Discher-Safran.2010.1,Trichet+Voituriez-Ladoux.2012.1} 
(\refsup{sup.othermodels}).  This is supported by our previous report
\cite{Mitrossilis+Asnacios.2010.1} that, when the external spring stiffness is
instantaneously changed in experiments, cells adapt their rate of force
build-up $\mathrm{d}{F}/\mathrm{d}{t}$ to the new conditions within $0.1$~s.
{This observation was repeated using an AFM-based technique
\cite{Webster+Fletcher.2011.1}.  In \cite{Crow+Fletcher.2012.1}, an overshoot
of the rate adaptation, which relaxed to a long-term rate within $10$ s, was
noted in addition to the initial instantaneous change of slope.} While this
instantaneity at the cell scale is not explained by mechanochemical regulation,
this behaviour is fully accounted by the mechanical model proposed here (see
\refsup{sup.one_d.stepchange}, \fig{fig.dFdt_vs_k},\subref{fig.change_k}).
Thus, the acto\-myosin cortex is mechanosensitive by essence: its peculiar
active visco\-elastic nature, which arises from collective effects, provides a
built-in system of adaptation to changes of the mechanical environment.

\setlength{\figtotwidth}{0.8\textwidth}
\setlength{\figwidth}{.38\figtotwidth}
\setlength{\figheight}{\figwidth}
\setlength{\margincrunch}{0\figtotwidth}
\addtolength{\margincrunch}{-0\linewidth}
\setlength{\figtotwidth}{.9\textwidth}
\begin{figure*}[htb]
\begin{tabular}{llll}
\hspace*{-\margincrunch}
\begin{subfigure}[b]{.5\figwidth}
 \centering\includegraphics[height=\figheight]{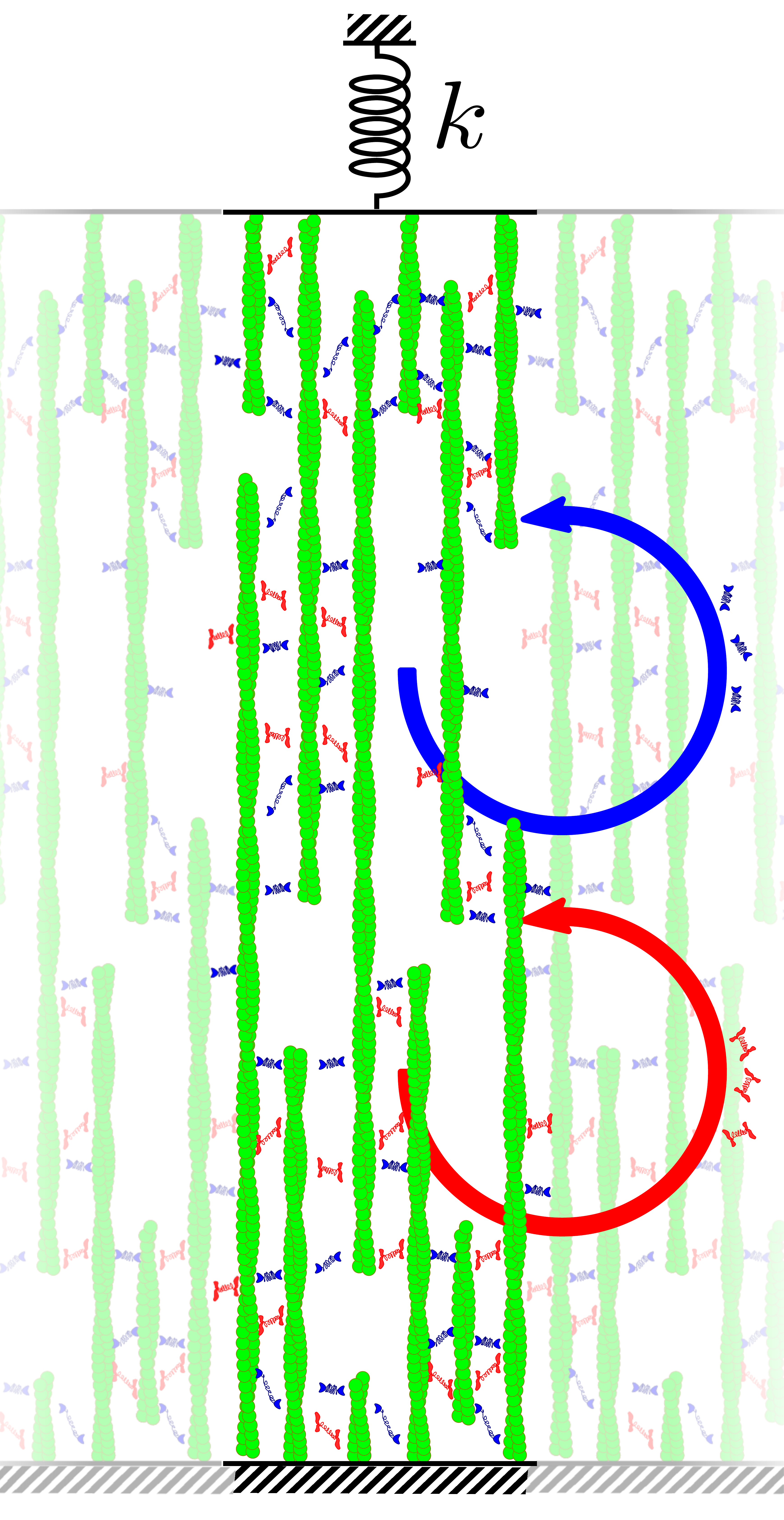}
\sublabel{fig.rheo.actin}
\end{subfigure}
&
\begin{subfigure}[b]{.48\figwidth}
\centering\includegraphics[height=\figheight]{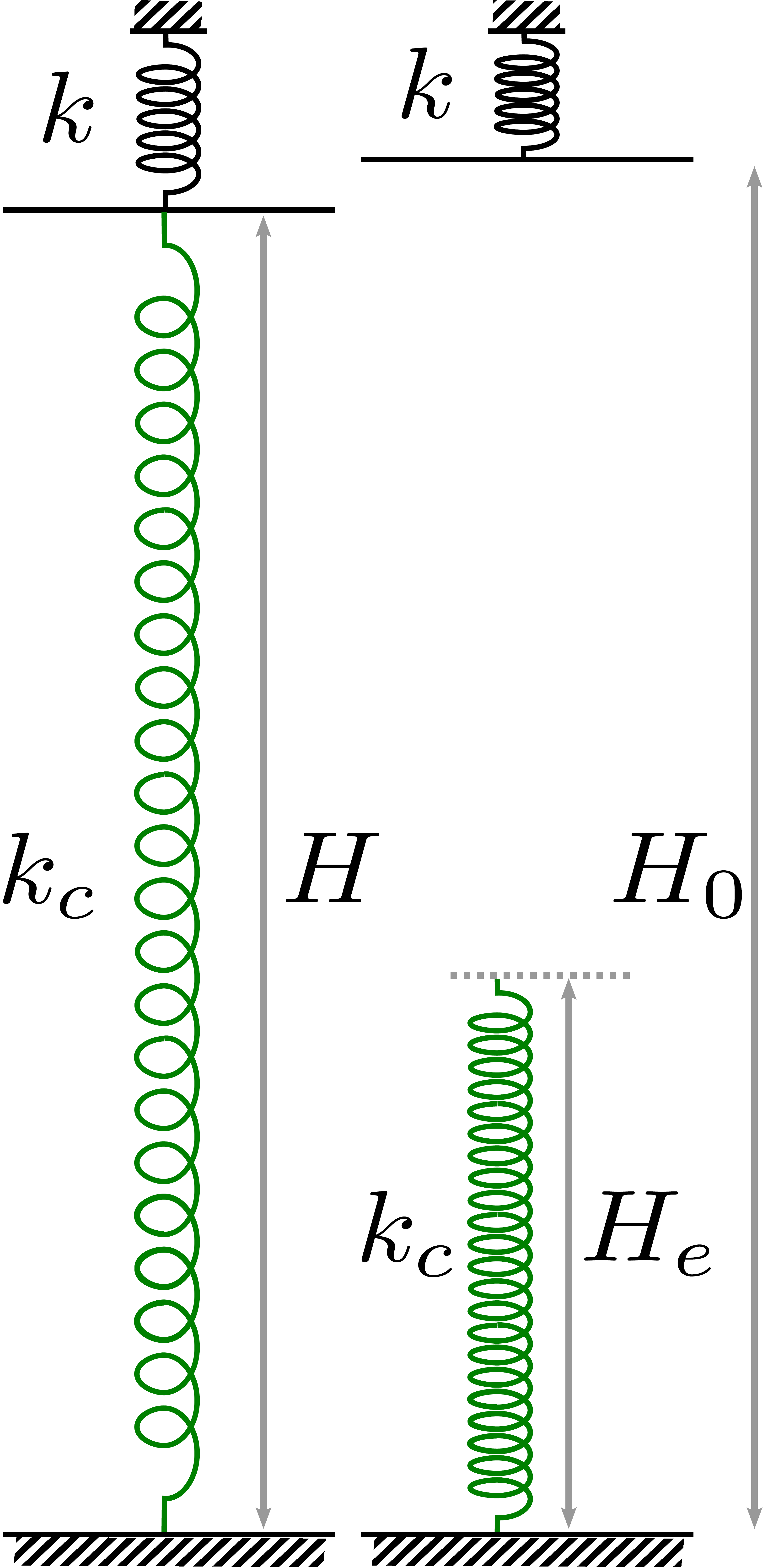}
\sublabel{fig.rheo.spring}
\end{subfigure}
&
\begin{subfigure}[b]{.95\figwidth}
 \centering\includegraphics[height=\figheight]{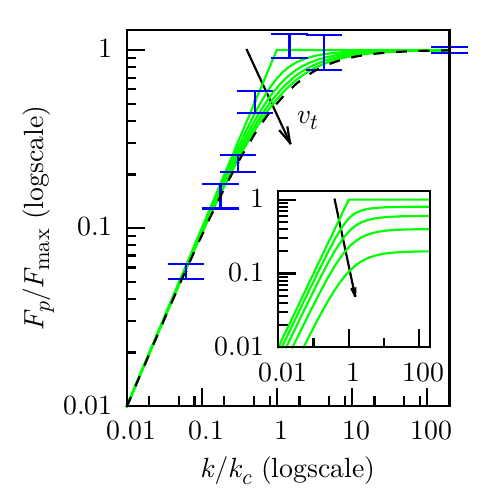}
\sublabel{fig.rheo.plot}
\end{subfigure}
&
\begin{subfigure}[b]{.7\figwidth}
 \centering\includegraphics[height=\figheight]{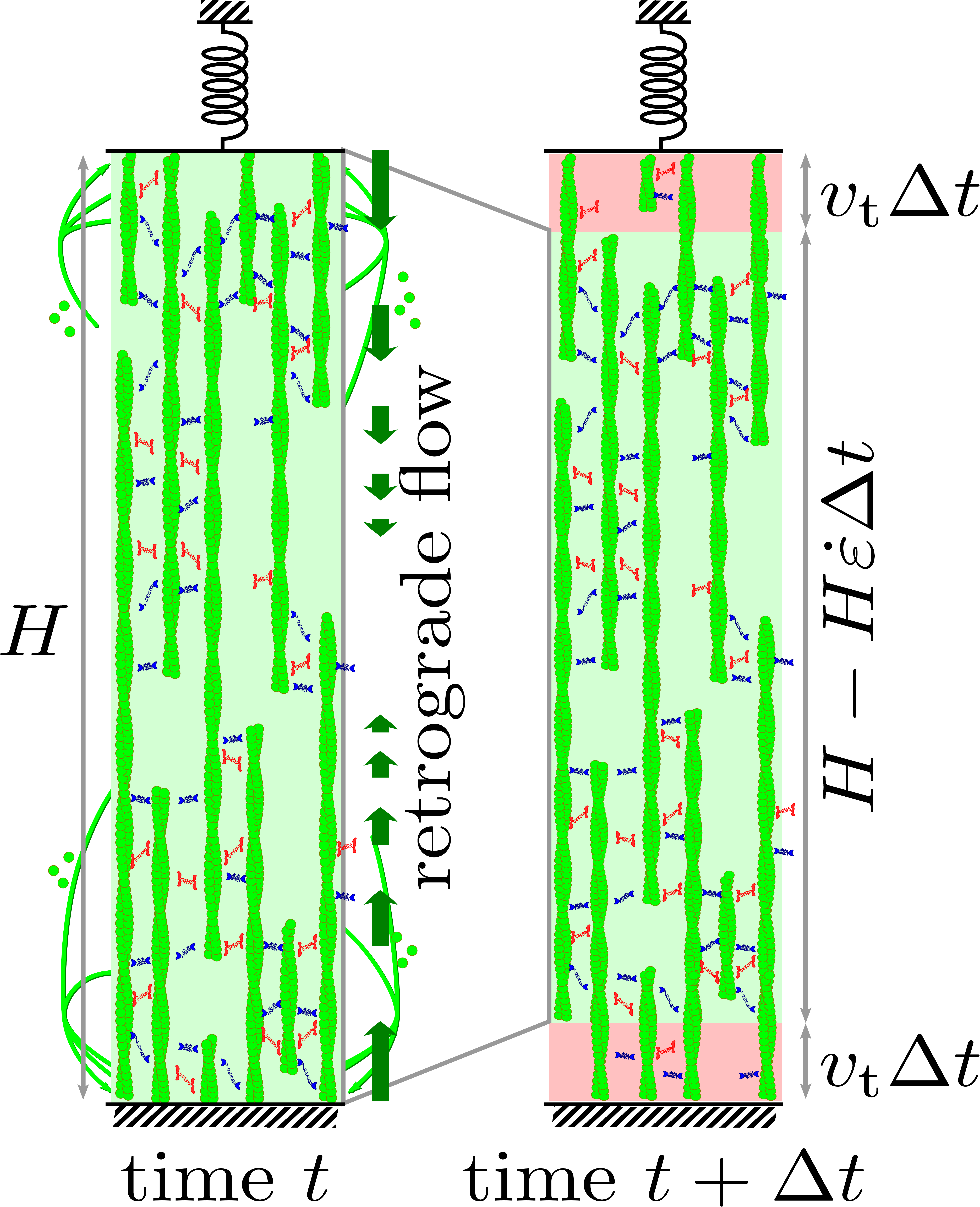}
\sublabel{fig.rheo.equilibrium}
\end{subfigure}
\hspace*{-.55\margincrunch}
\end{tabular}
\caption{A one-di\discretionary {-}{}{}men\discretionary {-}{}{}sion\discretionary {-}{}{}al simplification of the 3D-model preserves the essential mechanisms and results. (\emph{\subref {fig.rheo.actin}}) Sketch of the one-dimen\discretionary {-}{}{}sion\discretionary {-}{}{}al model, crosslinkers and myosin unbind and rebind so that the bulk properties are constant. (\emph{\subref {fig.rheo.spring}}) Spring model used elsewhere in the literature and to which the present model is compared. (\emph{\subref {fig.rheo.plot}}) Equilibrium state of models and cells as a function of microplate stiffness $k$, normalised by critical stiffness $k_c$ and maximum force $F_{\qopname \relax m{max}}$ in each case. Dashes: tension of the spring model, green: tension calculated in the 1D model for different values of the speed of treadmilling $v_{\mathrm {t}}$, blue: experimental results plotted with $k_c=\Sigma /(H_0+4\tau _{\alpha }v_{\mathrm {t}}) =41$ nN/$\mathrm {\upmu m}$ (no adjustment, see Supplementary Discussion \ref {sup.hill}). The stiffness-dependence of force in experiments is well-matched both by the spring model and the 1D model. Insert, tension calculated in the 1D model for different values of the speed of treadmilling $v_{\mathrm {t}}$, normalised by $F_{\qopname \relax m{max}}$ in the absence of treadmilling, $v_{\mathrm {t}}=0$: treadmilling reduces the force transmitted to the microplates. (\emph{\subref {fig.rheo.equilibrium}}) A dynamic equilibrium is attained when the retrograde flow exactly balances the speed of polymerisation at the boundaries, here sketched in the 1D model geometry.}
\label{fig.rheo}
\end{figure*}

\subsection*{Force-dependent regulation of cell size}
In microplate experiments, cells are observed to spread laterally simultaneously 
as they deflect microplates. These two processes both affect the arc distance between
the cell adhesions on each plate (\sfig{fig.shape}). Cell spreading is
known to be mediated by actin treadmilling 
\cite{Mitchison-Cramer.1996.1,Fournier+Verkhovsky.2010.1}, which controls the
extension of the lamellipodium \cite{Pollard-Blanchoin-Mullins.2000.1}. An effect
of treadmilling is that there is a net flow of filamentous actin from the lamellipodial region
into the proximal part of the actomyosin cytoskeleton, between adhesions 
\cite{Ponti+Danuser.2004.1,Fournier+Verkhovsky.2010.1}, which persists even if the 
cell and its adhesions are immobile \cite{Rossier+Sheetz.2010.1}. Thus the size
of the cell is regulated by the combination of the myosin-driven contraction
rate $-\deps$ of the cytoskeleton (the retrograde flow described above) and the
speed $\vp$ at which newly polymerised actin is incorporated into the cortex.
This feature can be included in the model as a boundary condition, prescribing a 
difference $\vp$ between the speed of the cell edge and the one of the actin cortex
close to the edge (\refsup{sup.one_d.treadmill} and \ref{sup.three_d}).
We find that this reduces the maximum tension that the acto\-myosin network can 
develop, however the shape of the dependence versus the external stiffness $k$ is little 
altered (\fig{fig.rheo.plot}). In particular, the force continues to be linearly dependent on $k$ 
for low stiffnesses, albeit with a reduced slope: this is a direct consequence of a mechanical 
regulation of cell height, which maintains
the microplate deflection $H_0-H_e$ when $k$ varies, thus $F=k(H_0-H_e)$ varies
linearly with $k$ in this range. 

Indeed, the
equilibrium height of the cell is obtained in the classical situation when the
polymerisation of actin at the cell edge is in balance with the retrograde flow
which drives the actomyosin away from the cell edge
(\fig{fig.rheo.equilibrium}). However, the interpretation here is not that
polymerisation generates this flow, but that myosin contraction is at its
origin, and that the equilibrium height is reached when the force balance
between myosin contraction and external forces acting on the cell is such that
retrograde flow exactly balances polymerisation speed.  
In the case when they do not balance, the cell edge will move at a speed which
is the difference between the speed of polymerisation and the retrograde flow
at the edge, until equilibrium is reached. In the case of our set-up, it is 
found that retrograde flow is initially faster than polymerisation, leading to
a decrease of cell height, and, because of the resistance of the microplates to
cell contraction, 
the tension $\sigma$ increases. In turn, this higher tension reduces the
retrograde flow until it is exactly equal and opposite to the polymerisation
speed.
Treadmilling and myosin
contraction thus work against one another, as has been noted for a long time
\cite{Mitchison-Cramer.1996.1} and is specifically described by Rossier et al.
\cite{Rossier+Sheetz.2010.1}. 

\setlength{\figwidth}{.37\figtotwidth}
\setlength{\figheight}{.75\figwidth}
\setlength{\margincrunch}{0\figtotwidth}
\addtolength{\margincrunch}{-\linewidth}
\begin{figure*}[tbp]
\begin{tabular}{lll}
\begin{subfigure}[b]{\figwidth}
 \includegraphics[height=\figheight,origin=tl]{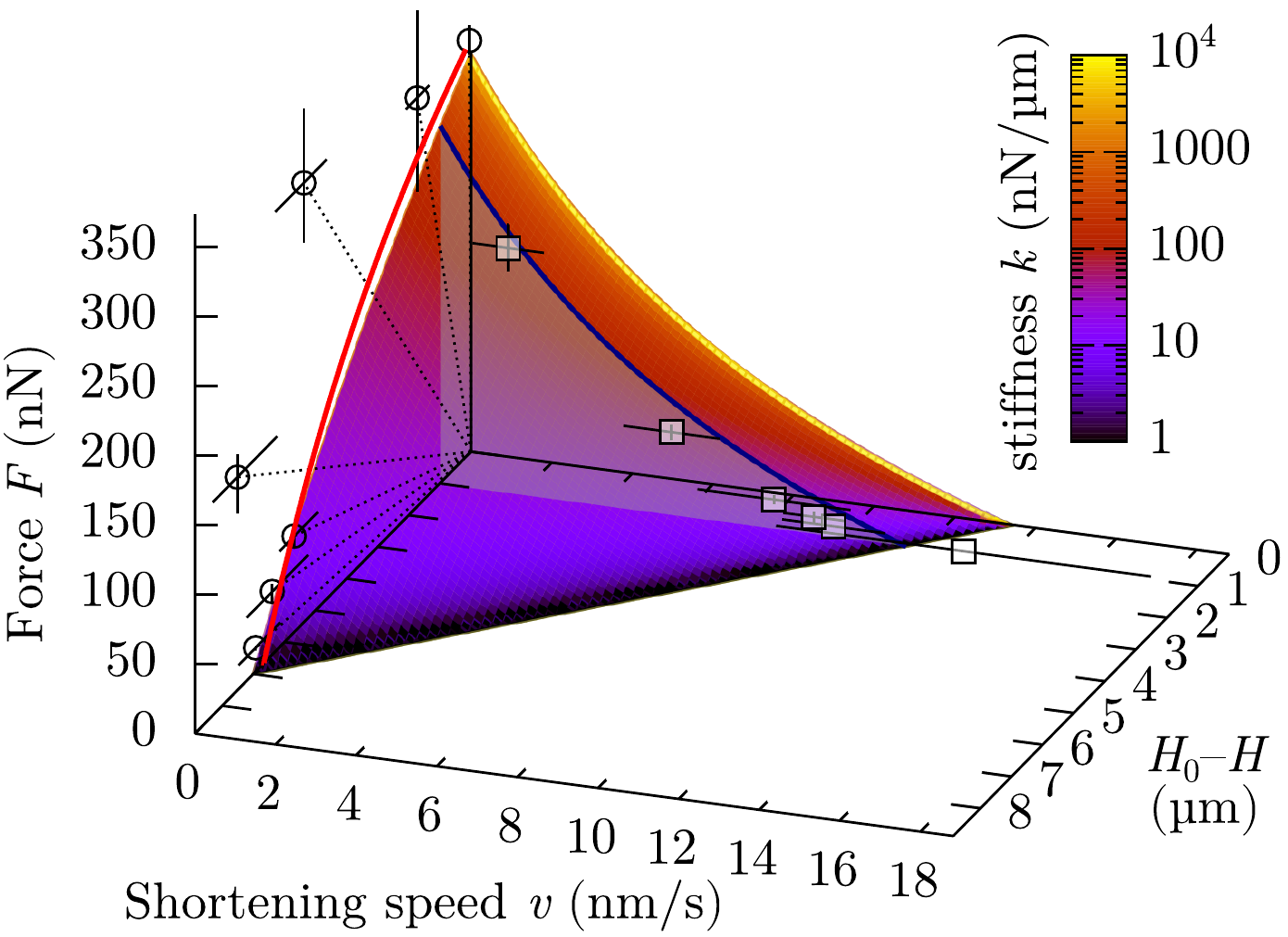}
\sublabel{fig.hill_values}
\end{subfigure}
&
\begin{subfigure}[b]{\figwidth}
 \centering
 \includegraphics[height=\figheight]{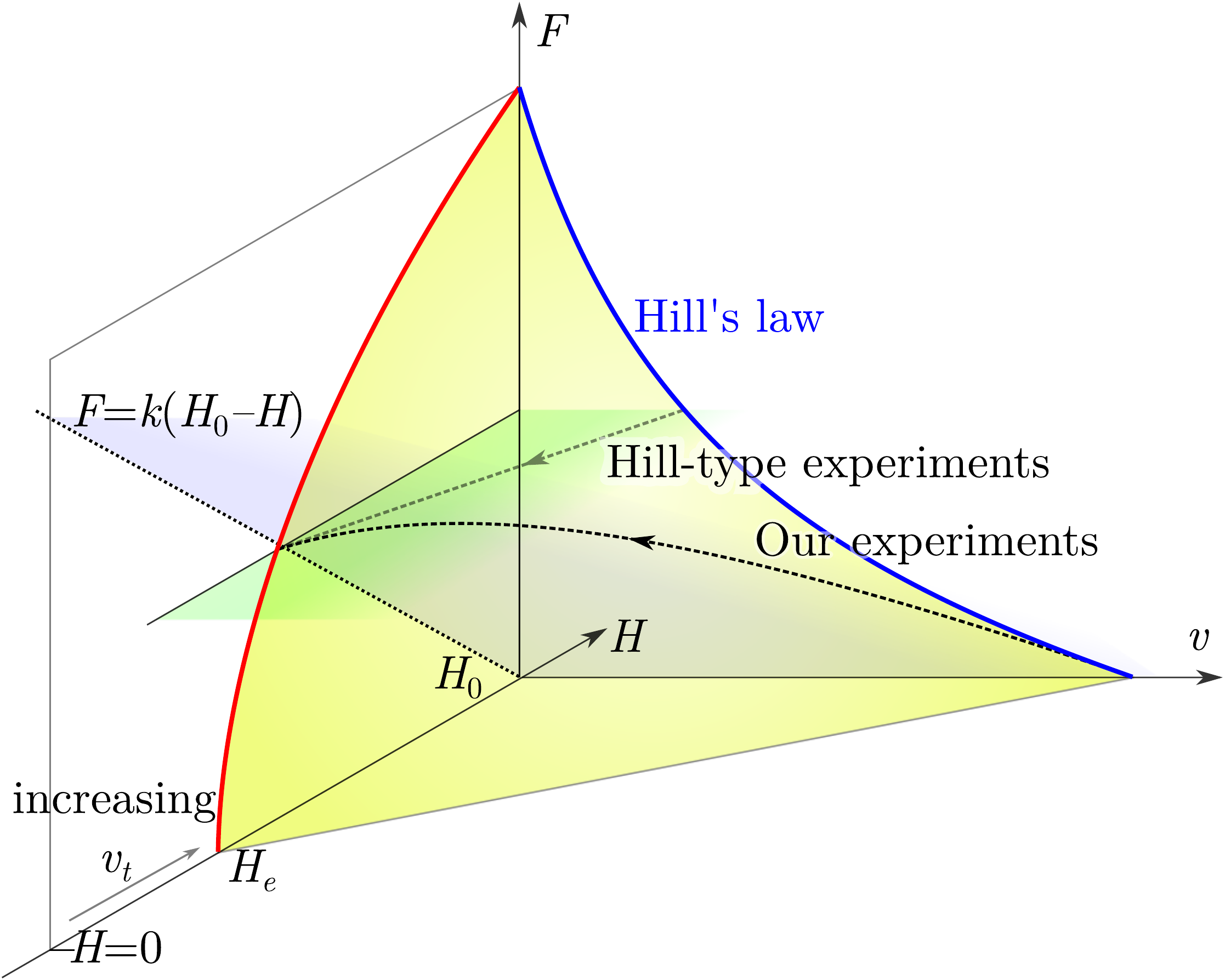}
\sublabel{fig.hill_sketch}
\end{subfigure}
&
\begin{subfigure}[b]{.7\figwidth}
 \centering
 \includegraphics[height=\figheight]{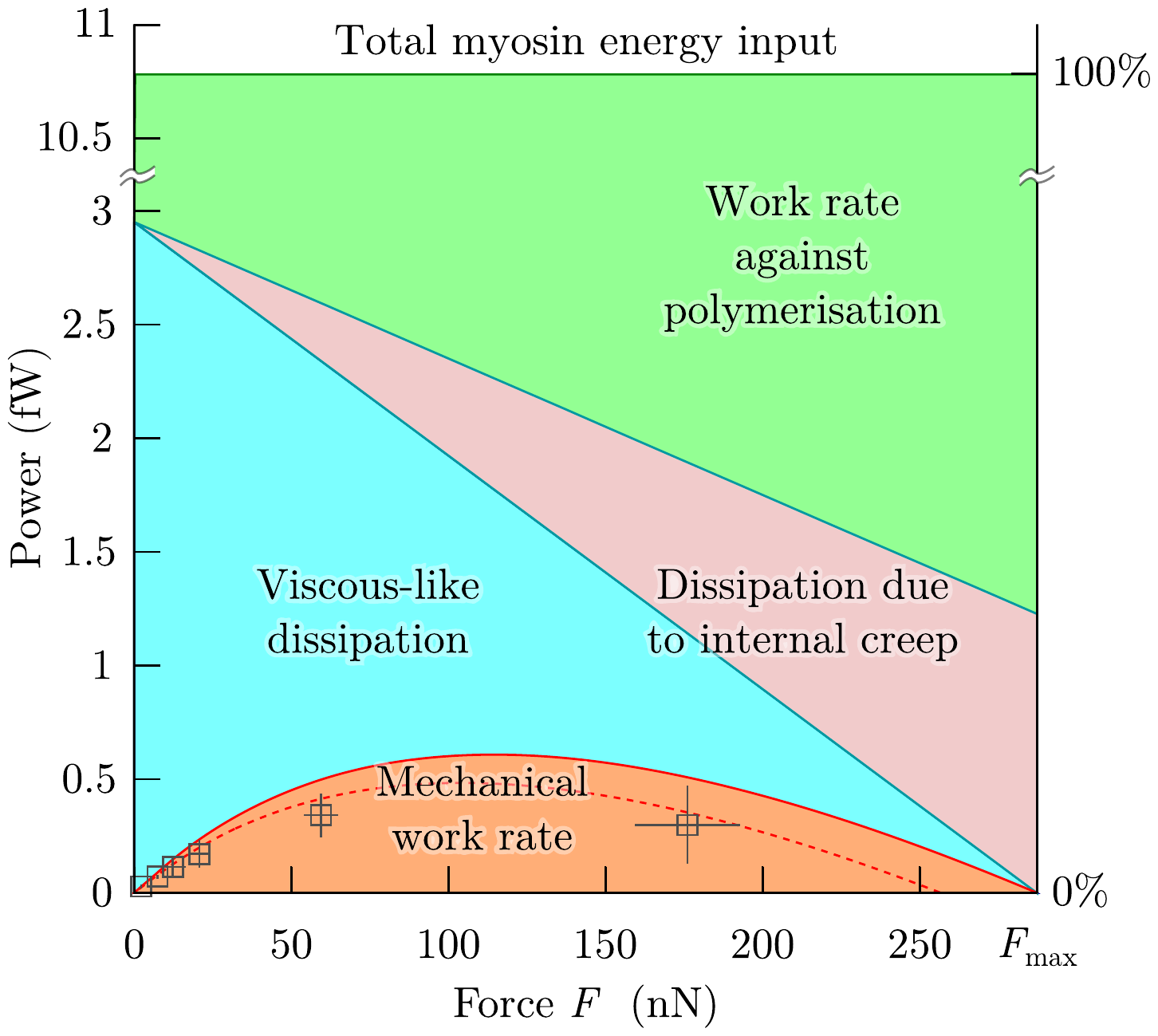}
\sublabel{fig.dissip}
\end{subfigure}
\end{tabular}
\caption{Liquid-like motor properties of cells. (\emph{\subref {fig.hill_values}}) The rheological model leads to a Hill-type equation (\ref {eq.hill}) which matches quantitatively the experimental data both during loading (blue curve at $H_0-H=1 \mathrm {\upmu m}$: force-velocity in 1D model, boxes: experiments) and at equilibrium (surface intercept with $v=0$: 1D model, red curve: 3D model, circles: experiments). Dotted lines correspond to the force-distance relationship imposed by a given microplate stiffness $k$. Same parameters used as in \fig{fig.data}. (\emph{\subref {fig.hill_sketch}}) Sketch of (\subref {fig.hill_values}). (\emph{\subref {fig.dissip}}) Power usage in a microplate experiment as a function of the external load $F$: only a small part of the load-independent myosin power is being transmitted to the cell environment as mechanical power, the rest is dissipated internally or compensates the antagonistic role of polymerisation. Boxes, experimental results, to be compared to the dashed line, mechanical power at $H_0-H=1\tmspace +\thinmuskip {.1667em} \mathrm {\upmu m}$, solid lines correspond to $H=H_0$. Same parameters used as in \fig {fig.data}.}
\label{fig.hill}
\end{figure*}

{These phenomena regulate cell size.}
For low external force, myosin-driven retrograde flow is high as the tension
that opposes it is small, and the balance between retrograde flow and
polymerisation speed is obtained when the cell has significantly reduced its
height, $H_e$ in \fig{fig.F_vs_k}. This height $H_e$ 
is thus a trade-off between 
the speed at which actin treadmilling produces new cortex $\vp$
(\refsup{sup.one_d})
and the rate 
of myosin-generated contractile strain $1/\Tc=\Sigma/(2\Ta E)=\alpha_{\text{myo}}(\ell\beta)^2/(2\tau_{\text{myo}})$, 
which is itself
the result of the frictional resistance of crosslinkers to myosin contraction.
Indeed, in the one-dimensional toy problem, the equilibrium height of the
system is $2 \Tc \vp$ for $k\ll k_c$, and thus the force developed by the external spring is
$F = k (H_0 - 2 \Tc \vp)$. In the three-dimensional full model of the cell cortex, this
equilibrium height is slightly modified as a function of its geometry, but is still proportional to
the product $\Tc \vp$ (\refsup{sup.three_d.equil}).
This equilibrium height is reached when the treadmilling speed balances exactly the speed at which
myosin in the bulk contracts the boundaries of the existing cortex, via a retrograde flow that involves 
the whole of the cortex but is maximum in distal regions (\fig{fig.rheo.equilibrium}).
This type of competition between
the protrusive contribution of actin polymerisation and the contractile contribution of myosin is of course
noted in crawling cells \cite{Small-Resch.2005.1}, it is
also observed in immobile cells where centripetal movement of actin monomers within filaments is noted
even when adhesive structures are limited to a fixed location on a micropattern \cite{Rossier+Sheetz.2010.1}. 
Thus, the cell-scale model and experiments allow to determine the speed of treadmilling, which is a 
molecular-scale quantity. We find $\vp = 6.5 \pm 1.5$ nm/s in the 1D model, and $4$ nm/s for
the 3D model, values which are in agreement with the literature \cite{Rossier+Sheetz.2010.1},
$4.3 \pm 1.2$ nm/s. The 1D model also allows to obtain the relaxation time of the crosslinked
actomyosin network, $\Ta=1186\pm 258$ s, consistent with elastic-like behaviour for frequencies higher
than $10^{-3}$ Hz, the contractile characteristic time $\Tc=521 \pm 57$ s, consistent with a $24$-minute
completion of actin super-precipitation \cite{Soares+Koenderink.2012.1} and 
$\Sigma S=(2.0\pm 0.9)\cdot 10^3$ nN, see \refsup{sup.one_d.quanti}. These values fit both
the plateau ($v=0$) force vs.\ stiffness experimental results, \fig{fig.F_vs_k} and the dynamics
of the experiments, \fig{fig.dFdt_vs_k}. Without further adjustment, they also allow to predict the dynamical 
adaptation of the loading rate of a cell between microplates of variable stiffness \cite{Mitrossilis+Asnacios.2010.1},
\fig{fig.change_k}, and to plot the force--velocity--height phase-portrait of the
experiments, \fig{fig.hill_values}.

From an energetic point of view, it may seem very inefficient to use up
energy  for these two active phenomena that counterbalance one another.
However, in a great number of physiological functions such as cytokinesis and motility,
either or both of actin polymerisation and myosin contraction are crucial. It is therefore highly
interesting that, combined together, they provide a spring-like behaviour to the cell while
preserving its fluid nature, endowing it with the same resilience to sudden mechanical aggression
as the passive mechanisms developed by some organisms, such as urinary-tract
bacteria \cite{Fallman+.2005.1} and insects \cite{Federle+.2001.1}.

\subsection*{Single cells have similar energetic expenses to muscles}
These antagonistic behaviours of polymerisation and myosin contractility entail
energy losses, which define a range of force and velocity over which the
acto\-myosin cytoskeleton is effective.  The study of the energetic efficiency
of animal muscle contraction was pioneered by A. V.  Hill \cite{Hill.1938.1},
who determined a law relating force $F$ and speed of shortening $v$:
$
(F + a)(v + b)=c,
$
where $a$, $b$ and $c$ are numerical values which depend on two values specific of a given
muscle, namely a maximum force and a maximum speed, and a universal empirical constant.
Hill's law was then explained using a model based on the muscle molecular structure 
by A. F. Huxley \cite{Huxley.1957.1}. Recently, we have shown that a
law of the same form describes the shortening and force generation of cells in the
present setup \cite{Mitrossilis+Asnacios.2009.1}. In particular, the maximum attainable force
and velocity are due to energy losses. 
The model can shed light on the molecular
origin of these losses, and leads to the quantitative force-velocity diagrams in 
\fig{fig.hill_values},\subref{fig.hill_sketch}. 
Indeed, 
in terms of $F$ and $v$, \eq{eq.cst}
yields (\refsup{sup.hill}):
\begin{minipage}[c]{\columnwidth}
\begin{align}
\left(\frac{F}{S} + E\right) \left(v + 2\vp + \va\right) 
 &= (\Sigma + E) \va - \frac{H\dot{F}}{2S}.
\label{eq.hill}
\end{align}
\vspace*{0.2\baselineskip}
\end{minipage}
Here $\va = H/(2\Ta)$ interprets an \emph{internal creep velocity}.
The right-hand side corresponds to the source of power (minus the internal
elastic energy storage term $H\dot{F}/2$), the left-hand side is the power
usage (up to a constant, $E \va$, added to both sides). The formal similarity
of this law with Hill's law for muscles is not a surprise when one compares the
present model with Huxley's model of striated muscle contraction.  Indeed, the
main components in both models are an elastic structure with transient
attachments and an ATP-fuelled `pre-stretch' of the basic elements of the
systems which, upon release, generates either tension or contraction, or a
combination of the two. Treadmilling $\vp$ in our model superimposes an effect
similar to the ones already present.  It is easiest to understand this law in
the extreme cases of zero speed or zero load, which correspond respectively to
maximum contractile force and maximum contraction speed.

The case of zero load, $F=0$, corresponds to the highest velocity. In the case of muscle
contraction, the fastest sliding of actin relative to myosin filaments is
limited in Huxley's model by the rate at which myosin heads detach after their stroke.
This is
because myosin heads which remain bound to actin will get entrained and exert an
opposing force to the motion. This transient resistance is similar to friction. In our
nonmuscle actomyosin model, crosslinkers also need to unbind so that the actomyosin
network, which is elastic at short times, fluidises and flows. The velocity it reaches 
is thus a decreasing function of the relaxation time $\Ta$.

Zero speed, $v=0$, corresponds to microplates of infinite stiffness. If in
addition protrusion via actin treadmilling is blocked, $\vp=0$, there is no net
deformation of the network (or sliding of the filaments in Huxley's model), 
however energy is still being dissipated: 
indeed, myosin motors will still perform power strokes and generate tension
in the network, but nearby crosslinkers and myosins themselves will also
detach at the rate $1/\Ta$. This detachment will result in the local loss of the
elastic energy that had been stored as tension of the network without resulting in 
a global deformation, corresponding to some \emph{internal creep}. This time, the
maximum force is an increasing function of relaxation time $\Ta$. 
In nonmuscle cells, actin treadmilling is still present when the cell edge is
immobile \cite{Rossier+Sheetz.2010.1}, and reduces the maximum force that can
be attained, because part of the myosin power will be used to contract this newly 
formed cortex.

In the intermediate regimes where neither $F$ nor $v$ are zero, the term $Fv$
corresponds to an actual mechanical work performed against the external load,
\fig{fig.dissip}. 
This work uses up the part of myosin energy that is not dissipated by internal
creep, by effective friction or by working against the actin-driven edge
protrusion.

\section*{Discussion}
The model described here is based on a simple description of collective
dynamics of actin and myosin that is consistent with 
observations at the protein scale \cite{Rossier+Sheetz.2010.1}, but does not include 
molecular sensitivity of the dynamics or actively driven reorganisation
of the actomyosin cortex. We find that the linear rheological law that
arises from this description allows to predict accurately the rigidity
sensing experiments that we carry over two decades of external rigidity.
The dynamics of cell pulling are also recovered, and we show that their
similarity with the dynamics of muscle contraction \cite{Hill.1938.1} are due 
to the parallelism that exists between actomyosin dynamics in nonmuscle cells
and the dynamics of thin and thick filaments in muscle \cite{Huxley.1957.1}.
Because the model is based on the collective dynamics of myosin and actin
filaments, it allows to understand their role in rigidity sensing: myosin
provides a contractile stress $\Sigma$ which will in turn generate traction 
forces at the cell--substrate contact area. In cases when $\Sigma$ is in 
excess to these traction forces (which resist cell contraction), a retrograde 
flow is generated, the cell contracts and deforms its environment.
This retrograde flow is limited by the time needed by the actin network to fluidise 
(viscoelastic relaxation time $\Ta$), and is force-dependent. Retrograde
flow also works against actin protrusion at the cell edge: we hypothesise that this
antagonism regulates cell shape, and show in the case of our parallel microplates
setup that this regulation of cell height determines the rigidity sensing
features of cells. The existence of a deformation set-point had already been
speculated on micropillar array experiments \cite{Saez+Ladoux.2005.1} and used
as a hypothesis in modelling work \cite{Zemel+Discher-Safran.2010.1}, here we
shed light on its relationship with molecular processes and retrograde flow.
Because the deformation set-point is obtained as a balance between protrusion and
contraction, it is versatile and is likely to be tuned by the many pathways known
to affect either of these. Undoubtedly, it is
a loss of energy for the cell to have such antagonistic mechanisms,
which we quantify in \fig{fig.dissip}. However, the total power of myosin action that we
calculate for a single cell is of the order of $10$ fW, which is less than one thousandth of the total power
involved in cell metabolism \cite{West+.2002.1}. It is thus not surprising that this energy expense is not optimised
in nonmuscle cells, as the evolutionary pressure on this cost can be deemed very low, while on the other hand the
same structure
confers to the cell its mechanical versatility and
reactivity to abrupt mechanical challenges. In this, the cell may be likened to a wrestler ready to face a sudden
struggle: the wrestler maintains a high muscle tone, having his own muscles work against one
another. Maintaining this muscular tone is an expense of energy, but the benefit of resisting assaults widely exceeds
this cost.



\section*{Materials and methods}
    \fontsize{8pt}{10pt}\selectfont
\subsubsection*{Cell culture, fibronectin coating and drugs}
Rat embryonic fibroblasts-52 (Ref52) line with YFP-paxillin, kindly provided by A. Bershadsky, Weizmann Institute,
and C2-7 myogenic cell line, a subclone of the C2 line derived from the skeletal muscle of adult CH3 mice, kindly
provided by D. Paulin and Z. Xue, Universit\'e Paris-Diderot, were cultured and prepared both using the procedure described 
in \citetext{Mitrossilis+Asnacios.2009.1}. 
Glass microplates and, in the case of the experiment shown in \sfig{fig.shape.TIRF}, the glass coverslip at the
bottom of the chamber were coated with 
$5 \upmu$g/mL as described in \citetext{Mitrossilis+Asnacios.2009.1}. 
In experiments described in \refsup{sup.microtubules} and \sfig{fig.microtubules}, 1 $\upmu$M Colchicin was used.

\subsubsection*{Side-view experimental procedure}
Ref52 or C2-7 cells were used in microplate experiments as described in \citetext{Mitrossilis+Asnacios.2010.1} with the
same equipment and reagents. Cells were then suspended in a temperature-controlled
manipulation
chamber filled with culture medium and fibronectin-coated microplates were brought in
contact with a single cell as described in \citetext{Mitrossilis+Asnacios.2009.1}. 
After a few seconds, the two microplates were simultaneously and smoothly
lifted to 60 $\micron$ from the chamber's bottom to get the desired configuration of
one cell adherent between two parallel plates. One of the plates was rigid, and
the other could be used as a nanonewton force sensor \cite{Desprats+Asnacios.2006.1}. 
By using flexible microplates of different stiffness values, we were
able to characterise the effect of rigidity on force generation up
to a stiffness of about $200$ nN/$\micron$. In order to measure forces at an
even higher stiffness, we used a flexible microplate of stiffness $\simeq 10$ nN/$\micron$
but controlled the plate--to--plate distance using a feedback loop, maintaining
it constant regardless of cell force and plate deflection 
 \cite{Desprats+Asnacios.2006.1,Mitrossilis+Asnacios.2010.1}.
Concurrently, we visualised cell spreading under brightlight illumination at an angle 
perpendicular to the plane defined by the main axis of the two microplates. 
For conditions referred to as `low stiffness', such experiments with $n=5$ Ref52 cells and $n=4$ C2-7 cells were analysed. 
For conditions referred to as `infinite stiffness', again $n=5$ Ref52 cells and $n=4$ C2-7 cells were analysed. In both cases,
the distribution of values obtained for the different cell types was not significantly different,
see \sfig{fig.compare_cell_types}.

\subsubsection*{Bottom-view experimental procedure}
Additionally, $n=4$ Ref52 cells were used in experiments using another experimental procedure allowing to
image the adhesion complexes at one of the plates. The objective of the TIRF-equipped microscope (Olympus IX-71,
with $488$ nm wavelength laser) was put in contact with the fibronectin-coated
rigid glass plate at the bottom of the chamber. A fibronectin-coated flexible
plate is put in contact with the cell after sedimentation. The deflection of
the flexible plate is measured by imaging its tip with a custom-made
microscope. The latter is composed of a computer-controlled light source, 
an optical tube (InfinitubeStandard, EdmundOptics), a long
working-distance objective (20X, Mitutoyo), a prism which orients the beam in
the direction of the flexible plate tip, and, eventually, a mirror
that reflects back light towards the prism. Thus the flexible plate tip is
imaged, through the objective and the optical tube, on a photo-sensitive
detector (S3931, Hamamatsu). A feedback procedure is applied as in
 \citetext{Mitrossilis+Asnacios.2009.1} in order to mimic an infinite stiffness
of the flexible microplate.

\subsubsection*{Image analysis and geometric reconstruction}
Images were treated with ImageJ software (National Institutes of Health,
Bethesda, MD). For side-view experiments, 6 geometrical points were identified at each
time position, corresponding to the 4 contact points of the cell surface with the microplates
and to the 2 extremities of the cell `equator', i.e. the mid-points where cell surface is perpendicular
to microplates. Assuming a symmetry of revolution, these points define uniquely
the cell equatorial radius $R_c$, the average radius at the plates $R_p$, the
cell half-height $h$, and the average curvature of the cell surface $\kappa$
(average of the inverse of the radii of the circles shown in
\sfig{fig.shape.circles}). For bottom-view experiments, only the radius $R_p$ at the bottom plate
can be acquired dynamically. The initial radius of cells when still spherical was measured
using transmission image microscopy, it was used as the value $h_0$ and was found to be consistent
with side-view measurements of $h_0$. In the case of infinite stiffness, we assumed that the
curvature of fully spread cells viewed from the bottom behaved as the curvature of side-viewed cells, 
$\kappa h_0 = 0.90 \pm 0.03$ ($n=9$). This allowed to estimate the fully-spread radius at the
equator, $R_c$, and to use $n=13$ experiments for identifying data in the fully
spread configuration for infinite stiffness.

\subsubsection*{Data analysis and model resolution}
Microplate deflection time series were converted to force time series as described in \citetext{Mitrossilis+Asnacios.2010.1}. 
Force data was then re-gridded onto the time positions of the microscopy images using the in-house open-source software DataMerge,
based on the LOESS implementation of the GNU Scientific Library. 
Analytical calculations were assisted by the open-source computer algebra system Maxima. Numerical simulations
of ordinary differential equations derived from the model were simulated using the GNU Octave scientific computing
programming language for figure \ref{fig.data}. Analytical implicit equations were plotted using
the open-source Gnuplot software in figures \ref{fig.F_vs_k}, \ref{fig.rheo.plot} and 
\ref{fig.hill_values}.


    \fontsize{8pt}{10pt}\selectfont

\subsubsection*{Acknowledgments}
J.E. thanks especially John Hinch, Claude Verdier, Martial Balland, Karin John, 
Philippe Marmottant and Cyril Picard for fruitful discussions.
J.E. wishes to acknowledge funding by R\'egion Rh\^one-Alpes (Complex systems
institute IXXI and Cible), ANR "Transmig" and Tec21 (Investissements d'Avenir - 
grant agreement n° ANR-11-LABX-0030). The experimental work
was supported in part by ANR funding (ANR-12-BSV5-0007-01,
"ImmunoMeca").





    \fontsize{8pt}{10pt}\selectfont

\end{multicols}








\clearpage
\def\usebibtex{0}
\def\showlegends{1}
\newcounter{savebib}
\setcounter{savebib}{54}
\begin{multicols}{2}
    \fontsize{9pt}{10pt}\selectfont

\begin{supplements}
\section*{\SIname}
\renewcommand{\thesubsection}{S\arabic{subsection}}                           
\setcounter{secnumdepth}{3}

\subsection{Role of microtubules}
\label{sup.microtubules}

It has been reported that in addition to the substrate, part (of order 13\%) of the cortical tension could
be balanced by the resistance to compression of microtubules \cite{Stamenovic+Wang.2002.1}.
We have thus controlled whether this was the case in our setup, in particular for low external
stiffness, which corresponds to lower height of cells and thus are geometrically more likely to
involve microtubule compression. The results, shown in \sfig{fig.microtubules}, indicate that there is
no such influence within experimental error. This allows us to neglect the role of microtubules compared
to actomyosin tension and microplate resistance to bending in the modelling that follows.

\subsection{Model derivation}
\label{sup.model}

As stated in the text of the article, we are looking for the simplest model consistent with
the fact that the actin plus crosslinkers network \emph{in vivo} is not able to resist extensional 
stress in the long term. This is consistent with a dominant loss modulus at low frequencies in
cell-scale rheological probing \cite{Wottawah+Kaes.2005.1}
and \emph{in vitro} studies \cite{Soares+Koenderink.2012.1}, and is linked
with the fact that crosslinkers \emph{in vivo} are
transient with a short residence time \cite{Fritzsche+Charras.2013.1}.
Basic models of transiently crosslinked networks based on rubber-like models
were first explored by Green and Tobolsky \cite{Green-Tobolsky.1946.1}
and Yamamoto \cite{Yamamoto.1956.1}, and their nonlinear properties are still being 
investigated \cite{Vaccaro-Marrucci.2000.1}. Their response, up to the first order,
turns out to be the same as the one of polymer solutions, that is, their
stress-strain relationship is governed by Maxwell constitutive \eq{eq.cst_nomyo},
$$
\Ta \ucm{\tsigma} + \tsigma - 2 \Ta E \dot{\teps} = 0,
$$
with
${\tsigma} = 2E \left(\beta^2\left\langle \ARE\ARE \right\rangle - \ts{I}\right)$ (\cite[p. 116]{Larson.1999.1},
{the parameter $\beta^2=3/(2N_K b_K^2)$ 
is related to $b_K$ the length and $N_K$ the number of Kuhn steps between two crosslinks}).
Here the upper convected Maxwell tensor derivative $\ucm{\tsigma} = \dot{\tsigma} 
- \grad\vc{v}\trsp \tsigma - \tsigma \grad\vc{v}$ takes into account the affine stretching of
the strand vectors  $\ARE$, the basic units of the network, by the velocity gradient $\grad\vc{v}$.
The difference is that, for polymer solutions, the time $\Ta$ is the ratio of solvent viscosity to
polymer elasticity, because this is the characteristic time at which the polymers can deform
relatively to an affine global deformation of their surroundings (the solvent,  \cite[p. 123]{Larson.1999.1}),
while, in the case of transiently crosslinked networks, $\Ta$ is the characteristic unbinding time
of the crosslinks. Thus, the product $\Ta E$ which has the dimension of a
viscosity is only some apparent viscosity at the macroscopic scale, and
corresponds in fact to an elastic energy dissipation at rate $1/\Ta$.
If there is a large number of crosslinkers present along a single filament, there will not be
a single relaxation time $\Ta$ but several \cite{Broedersz+MacKintosh.2010.1}. In the present work,
we choose to investigate the properties of the single-relaxation time model above because this allows
to calculate analytically the model solution while retaining the essence of a long-time viscous-like and
short-time elastic-like material.

A fraction $\alpha_{\text{myo}}$ of the crosslinkers considered are myosin bipolar filaments. In addition to their crosslinking role,
they may effectuate a power-stroke at a frequency $1/\tau_{\text{myo}}$, which results in ``sliding'' the corresponding
crosslinker by the myosin step length $\ell$. If $\psi(\vc{\rho})$ is the orientational distribution function \cite{Larson.1999.1},
this appears as additional sink and source terms in the right hand side of the probability balance equation,
\begin{align}
\dedt{\psi}& + \nabla_{\vc{\rho}}\cdot \left(\dot{\ARE}\psi\right) \nonumber
\\
	&= \frac{\alpha_{\text{myo}}}{\tau_{\text{myo}}} 
	\left( -\psi(\vc{\rho}) +\frac{1}{2}\psi(\vc{\rho}-\ell\vc{\rho}/|\vc{\rho}|) 
				+\frac{1}{2}\psi(\vc{\rho}+\ell\vc{\rho}/|\vc{\rho}|) \right)
				\nonumber
\\
	&\simeq \frac{\alpha_{\text{myo}} \ell^2}{2\tau_{\text{myo}}} \frac{\partial^2 \psi}{\partial |\vc{\rho}|^2}
	\label{eq.proba}
\end{align}
which, multiplied by $\vc{\rho}\vc{\rho}$ and integrated over all configurations, yields an additional term
of contractility 
$$
\Ta \ucm{\tsigma} + \tsigma - 2 \Ta E \dot{\teps} = \tSigma = \Sigma \ts{A},
$$
where $\Sigma = E \frac{\tau_\alpha}{\tau_{\text{myo}}}{\alpha_{\text{myo}} (\ell\beta)^2}$
is proportional to the myosin concentration and power-stroke frequency. The tensor $\ts{A}$
is the local orientation tensor of the actin fibres,
$$
\ts{A} = \int_{\vc{\rho}} \frac{\vc{\rho}\vc{\rho}}{|\vc{\rho}|^2} \psi\, \mathrm{d}\vc{\rho}.
$$
The ratio of the apparent viscosity $\Ta E$ and contractile stress $\Sigma$
provides us with another characteristic time, $\Tc = 2\Ta E/\Sigma = 
2\tau_{\text{myo}} \alpha_{\text{myo}}^{-1}(\ell\beta)^{-2}$, which
characterises the dynamics of shrinking of an acto\-myosin network in the absence
of crosslinkers, as is the case in \emph{in vitro} experiments
 \cite{Soares+Koenderink.2012.1}.

Note that both characteristic times of crosslinker unbinding $\Ta$ and of
myosin power-stroke $\tau_{\text{myo}}$ have been taken as constants, independent of
the stress or strain they are submitted to. It is of course well established that
there is a dependence of these parameters on stress and strain
\cite{Debold+.2005.1}, which would introduce a nonlinearity in the model. An
important, unsolved question is to determine whether these dependences are or
not major players in the cell-scale mechanical behaviour of actomyosin, e.g.
through a stress-driven ripping of crosslinks. As their effect is a nonlinear
variation of the above model, the standard modelling procedure is to study the
linear response first (with constant characteristics times), before considering
the full nonlinear model. It is found in the sequel that the linear model is
sufficient to reproduce the experimental data, that is, for our experiments,
collective effects explain the observed behaviour by themselves\footnote{This is
also the case for Huxley's model of muscle contraction \cite{Huxley.1957.1}, where
the author observes, p.~290, that
assuming specific force-dependent binding kinetics only tunes the system's efficiency,
not affecting its ability to fit Hill's force-velocity experimental observations.}. 
In these conditions, the effect of additional nonlinear terms could not be distinguished
from this linear baseline, and were therefore not introduced in the present
work.

The model obtained is thus the one of a viscoelastic liquid. This is in line with
the seminal model by He and Dembo \cite{He-Dembo.1997.1} who modelled the cytoskeleton
as a viscous fluid and used it in numerical simulations of the cytokinesis. This is
also similar to the actin dynamics part of the model used by 
 \citetext{Rubinstein+Verkhovsky-Mogilner.2009.1} to simulate keratocyte migration. 
We provide above a microstructure-based derivation of this class of models, which allows
us to interpret the dissipation in terms of molecular behaviours (see the discussion on
Hill's law, \refsup{sup.hill}). This type of model, to the best of our knowledge, was
never used to study mechanosensing behaviours of single cells. In the sequel
(\refsup{sup.othermodels}), we compare it to 
models that have been used to analyse cell-scale mechanosensing,
but will first investigate its basic predictions.

\subsection{One-dimensional problem}
\label{sup.one_d}

\subsubsection{In the absence of treadmilling}

In this section we investigate the behaviour of a material modelled by the
constitutive law, \eq{eq.cst}, in a simplified, one-dimensional geometry described
in \fig{fig.rheo.actin}. The acto\-myosin network is assumed to occupy an infinite
cuboid between two horizontal plates, one of which is fixed and the vertical
position of the other governed by a spring of stiffness $k/S$ (per unit area)
and equilibrium distance with the other plate $H_0$. If the current distance
between the plates is $H$, the force exerted by the top plate on the material is
thus $F/S = k(H_0-H)/S$ per unit area.

We assume that the bulk forces acting on the acto\-myosin network (such as
the friction with the cytosol) are negligible
compared to the force developed by myosin contraction, this writes:
\begin{align}
\div \tsigma = \vc{0},
\label{eq.momentum}
\end{align}
with the boundary condition $\tsigma \vc{e}_z = (F/S)\, \vc{e}_z$ at the upper
plate; and corresponds to neglecting friction with the cytosolic fluid in the
balance between the acto\-myosin stress and the force at the plates.
By symmetry, only the vertical component $\sigma_{zz}$ of the
stress tensor $\tsigma$ in the acto\-myosin material is nonzero, we denote it
$\zet$, from the above we have $\partial_z \zet=0$ and thus the boundary
condition imposes $\zet = F/S = k(H_0-H)/S$ at every $z$, the stress is fully transmitted
through the material. 

On the other hand, the rate-of-strain tensor $\dot{\teps}$ is also limited to a
vertical component $\dot{\eps} = \dot{H}/H$. Using these equalities, the
constitutive \eq{eq.cst} describes the complete dynamics of the system. In
particular, a permanent regime ($\dot{\eps}$=0, $\dot{\zet}=0$) is reached for
\begin{align}
F_p = \zet_p S &= \frac{\Sigma S}{2} 
        \left( 
                \frac{k+k_c}{k_c} 
                -\sqrt{ \left(\frac{k+k_c}{k_c}\right)^2 - \frac{4k}{k_c}  }
        \right)
	\nonumber
	\\
	&= \left\{\begin{array}{ll}
		k H_0 &\text{if }k<k_c,
		\\
		\Sigma S 	&\text{else},
	   \end{array}\right.
\label{eq.plateau.nopolym}
\end{align}
for $k<k_c$, with $k_c = \Sigma S/H_0$. The acto\-myosin model's response is thus
very close to the behaviour of a spring of stiffness $k_c$ and free length $0$, when
put in series with an external spring of stiffness $k$: Indeed, the asymptotes are the same for
$k\rightarrow 0$ and $k\rightarrow\infty$ (\fig{fig.rheo.plot}), which is not obvious since in the permanent regime
the model only includes viscous dissipation and contractility. We have thus
shown that \emph{a contractile fluid behaves like a spring} in these conditions.

Indeed, when the cell is pulling on a spring of stiffness $k$, the maximum
deflection it can impose to the spring is $H_0$, e. g. the initial cell height and,
consequently, the maximum height it can shorten. However, this maximum
deflection is achieved only if the maximum force generated by the cell
$\Sigma S$ is larger than the external spring force at maximum deflexion $k H_0$,
in other words only if $k$ is lower than $k_c=\Sigma S/H_0$. In that case, the
force reached is $k H_0$, thus proportional to $k$. In the other case $k>k_c$, the 
deflection is less than $H_0$ and is set by having an equal tension $\Sigma S$ in both
the external equivalent spring $k$ and the cell.

\subsubsection{In the presence of treadmilling}
\label{sup.one_d.treadmill}

We now introduce the fact that actin filaments \emph{in vivo} are constantly
being polymerised from one end (`plus' end) and depolymerised, mostly from the
other (`minus' end), which
results in the so-called treadmilling phenomenon \cite{Mitchison-Cramer.1996.1}. Assuming that this
treadmilling is at steady-state in the cell on average at the relevant time
scale of our experiment, the effect of treadmilling in the bulk does not affect
the modelling assumptions done in \refsup{sup.model}: the elastic modulus $E$ of the
crosslinked network at any instant will have a constant average. However, at the
boundary, some filaments will have their `plus' end oriented towards the
boundary and thus polymerisation of these will entail a net extension of the
material (before stress equilibration, depending on the boundary conditions).

In the framework of our one dimensional toy problem, this introduces a drift
between the deformation of the material and the displacement of the boundary,
which can be expressed as:
\begin{align}
\dot{H} = H \dot{\eps} + 2\vp
\label{eq.polym}
\end{align}
with $\vp = \ell_{\text{a}}/\tau_{\text{a}}$ the treadmilling speed, that is,
the extra length added by a monomer polymerising divided by the characteristic
time for an ATP-fuelled monomer addition. The factor $2$ is due to this effect
taking place at both plates.

When injected into the constitutive model, this modifies the level of force (and
height $H_p$) in the permanent regime:
\begin{align}
F_p = \zet_p S = \frac{\Sigma S}{2}
        \left( 
                \frac{k+k_c}{k_c} 
                -\sqrt{ \left(\frac{k+k_c}{k_c}\right)^2 - \frac{4k}{\Sigma S} (H_0-H_e)  }
        \right),
\label{eq.plateau.polym}
\end{align}
with a modified 
critical stiffness $k_c=\lfrac{\Sigma S}{H_0 + H_{\alpha}}$. Here
$H_{\alpha}=4\Ta\vp=4\Ta\ell_{\text{a}}/\tau_{\text{a}}$ is a characteristic
\emph{elastic length} of newly polymerised material. The critical stiffness
$k_c$ is lowered in proportion with the corresponding relative increase of height.
We have also introduced another new parameter, the length
$H_e = 2 \Tc \vp = 4 \frac{\tau_{\text{myo}}}{\Ta} \frac{\ell_{\text{a}}}{(\ell\beta)^2\alpha_{\text{myo}}}$.
This length is discussed in the main text, it is a trade-off between the rate at which actin
contracts under the effect of myosin, and the speed at which actin network expands by polymerisation
at its edges. It defines the shortest height that the cell achieves in the 1D model,
when the stiffness of the plates is vanishingly small: in that case, all of the work of
myosins is spent in contracting the part of actin network newly polymerised.

For a low stiffness, this expression has for asymptote $F_p  \sim  k(H_0-H_e)$,
and for a large stiffness, 
\begin{align}
F_p \rightarrow \Sigma S \frac{H_0-H_e}{H_0+H_{\alpha}}
\label{eq.plateau.Fmax}
\end{align}
This does not
change the qualitative spring-like response of the material, only the
equilibrium length of the equivalent spring is not zero anymore but $H_e$. This
means that as the device stiffness goes to zero, the height of the model
meshwork will tend to $H_e = 2 \Tc \vp$, which is a dynamic balance between the
speed $2\vp$ at which new material is added at the two boundaries (modelling
polymerisation), and the rate $1/\Tc$ at which the existing material is contracted
by the myosin motors. Inside this model material, even once the final length
$H_e$ is reached and the boundaries are immobile, there remains a continuous
centripetal flow (\fig{fig.atef.sketch}) that is very reminiscent of the
retrograde actin flow observed in both crawling \cite{Mitchison-Cramer.1996.1} and immobile spread
cells \cite{Rossier+Sheetz.2010.1}. Exactly as in these cases, this is made possible by
depolymerisation inside the model material. This is an additional
source of energy dissipation in steady state, when the cell is apparently at
equilibrium with constant length $H_e$ and force $F_p$.

\subsubsection{Dynamics}
\label{sup.one_d.dyn}

The above constitutive and force balance equations can be written in terms of one equation
only with the tension as the unknown,
\begin{align}
\ded{\zet}{t} 
 = \frac{
	\Sigma\left( k(H_0-H_e)/S - \zet\right) 
	+\zet\left( \zet - k(H_0+H_{\alpha}) \right)
	}{
	\Tc\Sigma + \Ta(\zet+k H_0/S)
	}.
\label{eq.dzetdt}
\end{align}
One can analyse the rate of tension increase
when the system begins to pull and $\zet$ is still much smaller than $\Sigma$, for a low 
stiffness $k \ll ES/H_0$, 
\begin{align}
\ded{F}{t} = \frac{k(H_0-H_e)-F(t)}{\Tc} \sim \frac{F_p(k)-F(t)}{\Tc}.
\label{eq.one_d.Tc}
\end{align}
And for high stiffness $k \gg ES/H_0$,
\begin{align}
\ded{F}{t} &= \frac{\Sigma S (H_0-H_e) - F(t)(H_0+H_{\alpha)}}{\Ta H_0} \\
	&= \frac{(F_p(k)-F(t)) \, (H_0+H_{\alpha})}{\Ta\, H_0}.
\label{eq.one_d.Ta}
\end{align}

Experiments for very low values of the microplate stiffness $k$ allow to identify the parameters
$\Tc$ and $\vp$ of the model.

From $n=9$ experiments with $k \leq 1.6$ nN/$\micron$,
we find $2\Tc\vp = H_e = 6.6 \pm 1.1 \,\micron$ and $H_0=13.2\pm 0.1 \,\micron$. 
This is consistent with the fact that $\lfrac{\partial F}{\partial k} = H_0-H_e = 6.8\pm 0.4 \,\micron$
in \citetext{Mitrossilis+Asnacios.2009.1}.
Additionally, the maximum force attained $F_p$, the current force $F(t)$ and
the rate of growth of this force $\lded{F}{t}$ allow to calculate $\Tc$ using \eq{eq.one_d.Tc},
$\Tc = (F_p-F(t))(\lded{F}{t})^{-1} = 521 \pm 57$ s. This yields $\vp = 6.5 \pm 1.5$ nm/s, which is 
consistent with the literature \cite{Rossier+Sheetz.2010.1} as stated in the main text.

\subsubsection{Response to a step-change of external stiffness}
\label{sup.one_d.stepchange}

In \citetext{Mitrossilis+Asnacios.2010.1}, we are able to vary instantaneously (within $0.1$ s) the
stiffness $k$ of the external spring, while ensuring that there is no instantaneous change of the
force $F$ felt by the cell or of the microplate spacing $H$. In the modelling, this corresponds to 
an instantaneous change of both $k$ and $H_0$ at time $t^*$ such that $F$ and $H$ are continuous.
We can thus write the following relations, ensured by the experimental setup:
$$
k(t) = \left\{
  \begin{array}{l}
  \!k_0\text{, } t<t^*,
  \\
  \!k_1\text{, } t>t^*,
  \end{array}\right.
$$
$$
\begin{array}{ccc}
\!\!\!F(t)=F^* + \phi(t),
&
\!\!H(t)=H^* - \frac{\phi(t)}{k(t)},
\end{array}
$$
where $\phi(t)$ is the variation of force around the force at time $t^*$, $\phi(t^*)=0$.

Changing only the stiffness in this manner ensures that the cell does not feel any step-change: 
indeed, force and geometry are preserved, the only change is the \emph{response of the microdevice to 
a variation of the force applied to it}.
The experimental result is that, despite the fact that none of the physical observables have been
changed for the cell, its rate of loading $\lded{F}{t}$ of the device is instantaneously modified. In  
 \citetext{Mitrossilis+Asnacios.2010.1}, it is found that, when going from a stiffness $k_0$ to a stiffness $k_1$,
the new rate of loading matches the rate of loading that the same cell type exhibits in a constant stiffness $k=k_1$
experiment, at a time such that $F=F^*$. In \citetext{Crow+Fletcher.2012.1}, this experiment is repeated
with another cell type and it is also found that the rate of loading is instantaneously changed.
However, the value $\lded{F}{t}$ for $t\gtrsim t^*$ exhibits an overshoot: it
is initially different from the value of the corresponding 
constant stiffness experiment, and exhibits a relaxation towards it. 

We can rewrite \eq{eq.dzetdt} around $\zeta^* = F^*/S$, we find:
\begin{align}
\ded{F}{t}(t^*) = k(t) 
	\frac{H^*(\Sigma S - F^*) - H_{\alpha} F^* - H_e \Sigma S}
	     {\Tc \Sigma S + \Ta ( k(t) H^* + 2F^* ) }.
\label{eq.dFdt}
\end{align}
The result is thus a step-function, whose value for $t\gtrsim t^*$ is exactly the rate of growth
predicted by the model for a constant stiffness experiment with $k=k_1$. This corresponds to
the experimental results in both  \citetext{Mitrossilis+Asnacios.2010.1} and
 \citetext{Crow+Fletcher.2012.1},
except for the overshoot found in the latter. 

The step-change of the rate of loading can thus be explained by a purely intrinsic property of the
cell's cytoskeleton, as developed in the main text.

In  \citetext{Crow+Fletcher.2012.1}, a model is proposed that accounts for a step-change and
relaxation, as they observe experimentally. However this model has a limited validity
around $t^*$ and breaks down at long times, predicting infinite force. Here, our model
originally aiming at describing the long-time behaviour of the cell depending on stiffness
predicts the main feature of the experiments (the step-change of the rate of loading), but
not the overshoot found in one of the experiments. Our model however may produce this
type of overshoot if more than a single relaxation time is used. This would correspond to 
replacing the spring $k_1$ in the model in  \citetext{Crow+Fletcher.2012.1} by our
visco\-elastic constitutive equation.
This is found not to be necessary to explain the data studied here.

In \fig{fig.change_k}, we give a numerical simulation result that corroborates
the step-change of $\lded{F}{t}$ found in \eq{eq.dFdt} and can be compared to 
an experimental curve without any parameter adjustment. The time-profile of the 
microplate effective stiffness is imposed to be the same in the numerical simulation
as in the experiment. The profile of force increase obtained presents the same
instantaneous change of slope as the stiffness is varied, and the overall profiles
match quantitatively. 

\subsubsection{Hill's law}
\label{sup.hill}

We examine the power dissipation predicted by the one-dimensional model 
between two boundaries. The boundaries move towards the sample
at velocity $v$ (which is thus positive when the sample contracts) and feel a force $F$
exerted by the sample (positive in the direction of contraction). 

Hill's historical experiment \cite{Hill.1938.1} takes place at a constant level of force
$F$. Our experiments, on the other hand, prescribe a relationship between $F$ and
$v$ through the microplate stiffness, $\lded{F}{t} = -kv$. Both can scan the $F$--$v$
relationship by varying $F$ for the former or $k$ for the latter (in the case of
 \cite{Mitrossilis+Asnacios.2009.1}, we needed to investigate this relationship for a fixed
$H=H_0-\delta$ where $\delta$ is small), \fig{fig.hill_values} and \subref{fig.hill_sketch}. The calculation below does not
require to use either of these experimental ways to scan the  $F$--$v$
relationship. It relies only on the material properties of the sample.

As above, the mechanical equilibrium of the sample imposes that the vertical component of
the stress tensor $\zet$ is equal to $F/S$. Also, the treadmilling produces a mismatch
between the plate velocity $v=\dot{H}$ and the recoil of the existing network at the plate
$-H\deps$, expressed by $v=-H\deps+2\vp$ (see \eq{eq.polym}). These relations can be
injected into the constitutive \eq{eq.cst}:
$$
\Ta\left(\dot{F} - 2\deps F\right) + F - 2\Ta E S \deps = \Sigma S.
$$
We recover a virtual work formulation by
multiplying this by the velocity $H/(2\Ta)$. Adding the constant $ESH/(2\Ta)$ to both sides, 
this work can now be factorised in the manner of Hill's law:
\begin{align}
\left(\frac{F}{S} + E\right) \left(v + 2\vp + \frac{H}{2\Ta}\right) 
 &= (\Sigma + E) \frac{H}{2\Ta} - \frac{H}{2S}\dot{F}.
\label{eq.hill_jo}
\end{align} 
The meaning of some of these terms is explained in the main text.

In the case when the velocity $v$ is zero, 
the force generated is finite as calculated above, \eq{eq.plateau.Fmax}, which can also write:
$$
\frac{F_{\max}}{S} = \Sigma \left( 1 - \frac{E+\Sigma}{\Sigma} \frac{2 \vp}{\va + 2\vp}\right)
$$
Even if both $v$ and $\vp$ are zero, and thus the acto\-myosin does not contract macroscopically ($\deps=0$), the 
force generated remains finite since it does work at a molecular scale.
Actin polymerisation, by adding new material at the edges, introduces a `boundary creep', which
consumes additional work in conditions of fixed length ($v=0$).

As stated in the main text, apart from the polymerisation `boundary creep', this is the same 
dissipative mechanism as in the model of
muscle contraction by Huxley \cite{Huxley.1957.1}. In this model, myosin `elastic tails'
are prestretched preferentially in one direction before binding the actin thin filament.
When there is no net sliding ($v=0$), they eventually unbind without having had the 
opportunity to provide work, that is, they have conserved this level of stretching. Although
this is not explicitly written in the 1957 paper, the prestretch that had been bestowed
on the myosin `elastic tail' is thus lost.

\smallskip

Zero force $F=0$ condition is a (theoretical) limit corresponding to zero-stiffness of
the microplate. This is not attainable experimentally, as cells do not spread on both
plates if their displacement does not generate any external force---which corresponds to
the impossibility to apply any normal force to the plates. This effect probably has to do
with the mechanism of force reinforcement of adhesions. In the model however,
the limit can be studied, and yields a maximum velocity,
\begin{align*}
v_{\max} + 2\vp = \frac{\Sigma \va}{E} = \frac{H}{\Tc}.
\end{align*}
This corresponds to the power injected by the myosin motors, divided by the elastic
modulus of the crosslinked network (minus the treadmilling contribution): indeed, in this
limit, the myosins work against the elasticity of the acto\-myosin itself. The maximum
velocity is thus limited by the rate $\Ta$ at which the actin network fluidises thanks to
the detachment of crosslinkers. In comparison to the case of muscles, actin treadmilling reduces
the maximum speed of shortening by $2\vp$, as the receding speed of the edge of the actin
network must compensate for this speed of protrusion.
Quantitatively, using the values obtained in \refsup{sup.one_d.dyn}, we predict 
$v_{\max} \simeq 12.3$ nm/s, this is very close to $v_{\max}=13$ nm/s published in 
 \citetext{Mitrossilis+Asnacios.2009.1}.

Again, the dissipation here is of the same nature as for muscles in the model by Huxley: in his
case, the only crosslinker between thin and thick filaments are myosin heads, and
zero force is obtained at the finite velocity at which the work rate performed by pulling
myosin heads is exactly balanced by the work rate needed to deform myosin heads that
have not yet detached from the actin filament. They will eventually detach, thus dissipating
at a fixed rate the elastic energy that has been transferred to them.

\smallskip

When neither $F$ nor $v$ are zero, there is a nonzero productive work performed on the plates. Because of the existence of 
maximum force and velocity, it is necessarily written 
$$
\frac{Fv}{F_{\max} v_{\max}} = r\left(1 - \frac{F}{F_{\max}}  - \frac{v}{v_{\max}}\right). 
$$
In both the experiments by A. V. Hill \cite{Hill.1938.1} and ours \cite{Mitrossilis+Asnacios.2009.1},
$r$ is found to be mostly independent of
$F$ and $v$, and $r \simeq 0.25$. In Huxley's model of muscles, $r$ is a
signature of the preferential pre-stretch of myosins that models the power-stroke,
normalised by the detachment rate \cite{Huxley.1957.1,Williams.2011.1}. 
In our model of cells, $2r \simeq \Tc/\Ta$ is the ratio of the characteristic
times of contraction and of stress relaxation through crosslinker
unbinding. This does not explain the coincidence of finding the same value of $r$ in both
cells and muscles, however we note that this parameter has a similar signification in both
models.

Specifically,
\begin{align}
r
 &= \frac{E}{\Sigma\left( 1 - \frac{E+\Sigma}{\Sigma}\frac{2\vp}{\va+2\vp}  \right)}.
\end{align}
Thus in our case, Hill's parameter $r$ is such that
\begin{align}
\frac{E}{\Sigma} = \frac{\Tc}{2\Ta} \leq r \leq \frac{\Tc}{2\Ta} + \left(\frac{\Tc}{2\Ta}\right)^2,
\label{eq.one_d.r}
\end{align}
depending on $\vp$. 

Note that these equations can be applied to the acto\-myosin pushing against an obstacle, with 
$0\leq -F \leq ES$ and $0 \leq -v \leq \vp$. 

\subsubsection{Quantitative analysis}
\label{sup.one_d.quanti}

Since $r \simeq 0.25$ in experiments \cite{Mitrossilis+Asnacios.2009.1},
${\Tc}/{\Ta}$ has to be in the range $0.4$ to $0.5$, thus since $\Tc=521
\pm 57$ s, we have  $\Ta = 1186 \pm 258$ s.  In turn, using also $\vp=6.5$
nm/s (see \ref{sup.one_d.dyn}), we find $\Sigma S = F_{\max}\lfrac{(H_0+H_{\alpha})}{H_0-H_e} = (2.0\pm 0.9)\cdot 10^3$ 
nN ($n=13$). Thus the four parameters of the 1D model, namely $\Ta$, $\Tc$, $\vp$ and
$\Sigma S$ are identified using only the average maximum force (equilibrium of infinite stiffness 
experiments), maximum velocity (dynamics of experiments with very low stiffness), and
shortest height at equilibrium (experiments with very low stiffness), plus the shape of the
Hill-type law (parameter $r$). 

The other experimental data (stiffness-dependence of the force,
\fig{fig.rheo.plot}; dynamics as the microplate stiffness is modified, \fig{fig.change_k};
and values of $F(t)$ at $H_0-H=1 \micron$, \fig{fig.hill_values})
are matched by the model without any adjustable parameter. E.g.,
the values found yield a critical stiffness $k_c=\Sigma S/(H_0+4\Ta\vp) =41$
nN/$\micron$, which is consistent with the experimental results,
\fig{fig.rheo.plot}.

Moreover, the values for $\Tc$, $\Ta$ and especially $\vp$ are independently measurable and
are consistent with the literature, see text.

\subsection{Comparison with other models featuring cell-scale mechanosensing}
\label{sup.othermodels}

In \citetext{Zemel+Discher-Safran.2010.1,Zemel+Discher-Safran.2010.2}, the cell is
modelled as a linear elastic body having an intrinsic equilibrium shape, and
which is prestretched to some maximum strain, in our notation:
$$
\tsigma = 2E (\teps - \teps_0),
$$
with $\teps_0$ an equilibrium shape that the cell would take in the absence of external
stresses.
They supplement this model with 
a phenomenological feedback on the elastic modulus, on the time scale of hours or
days, which corresponds to the phenomenon of stress-fibre polarisation 
 \cite{Curtis+.2006.1}. This long term effect is out of the scope of our model and
experiments, thus our model compares to theirs before this feedback comes
into play. In a one-dimensional setting, their model thus corresponds to the
prestretched spring in \fig{fig.rheo.spring}, which yields a stiffness-dependence
closely mimicking experimental data and our model predictions, \fig{fig.rheo.plot}.
Our model in addition provides a microstructure basis for this qualitative behaviour
and a quantitative reading of experiments, in particular the equilibrium shape
$\teps_0$ and cell stiffness $k_c$ in \citetext{Zemel+Discher-Safran.2010.1} are explicited,
respectively, as proportional to the product of a characteristic contraction
time and the treadmilling speed, $\Tc\vp$, and as the ratio of the myosin
contractile stress and a length, $k_c=\Sigma S/(H_0+4\Ta\vp)$.

In \citetext{Trichet+Voituriez-Ladoux.2012.1}, they introduce a model also in the framework 
of active gel theory. However, they resort to the stress-strain relationship of
 \citetext{Zemel+Discher-Safran.2010.1}
so as to avoid to calculate the orientation tensor $Q_{ij}$ of
the cytoskeleton and use a 1D linear elastic stress-strain relationship, their
equation (S7) writes in our notation $\sigma = 2E(H-H_e)$.

In \citetext{Marcq+Prost.2011.1}, they present a 1D visco-elastic solid model, combining their
equations (2) and (5) writes in our notation:
$$
\sigma =  2E (H - H_e) + 2\eta \dot{H}.
$$
with $H_e = l^0_C - F_S/k_C$ in their notation.
To the difference to the previous models, the viscous term added in this model allows to
study the dynamics of the cell shortening. However, the equilibrium shape in all three models
is a static elastic balance between the environment resistance to deformation and a 
phenomenological internal elasticity, which corresponds to the spring model described in
\fig{fig.rheo.spring}.

\subsection{Three dimensional problem}
\label{sup.three_d}

In this section, we present a full three-dimensional model of the cell mechanics as a 
contractile visco-elastic thin shell, obeying the rheological \eq{eq.cst}, and enclosing
an incompressible cytosol. Forces are transmitted to the microplates at the contact line
between the cell boundaries and the microplates.

\subsubsection{Geometry}
\label{sup.geometry}

It is seen from experimental observations that the cell boundaries connecting the plates
are in most cases very well approximated by an arc of circle (\fig{fig.atef.sketch}, insert). 
This had already been shown in the case of cells spread on a microneedle array
having reached a stationary shape \cite{Bischofs+Schwarz.2008.1}, in the present setup 
we find that it is also true while the cell is spreading (\sfig{fig.shape.circles}). 

When observing cells from the side, we assume that the cell shape is cylindrical. This is 
supported by other experiments where cells are observed from the bottom in TIRF, \sfig{fig.shape.TIRF}.
Thus, along the axis $z$ orthogonal to the microplates, whose location is parametrised as $z=\pm h$ where 
$h$ is the half-height of the cell, we can fit experimental results using the law:
\begin{align}
r(z) = R_c + \frac{1}{\kappa}\left(1-\sqrt{1-(\kappa z)^2}\right), 
\label{eq.arccircle}
\end{align}
where $R_c$ is the radius
at the cell equator ($z=0$) and $\kappa$ the signed curvature in the vertical plane, \fig{fig.atef.sketch}.
The curvature $\kappa$ evolves in time from a positive curvature ($t = 10$ s in \sfig{fig.shape.circles}) to
a negative one ($t = 120$ s and later in \sfig{fig.shape.circles}). 

To the physicist it may be a surprise that the cell shape is not well fitted by a minimal surface such as a catenoid.
Indeed, although each boundary seen on \sfig{fig.shape.circles} can be reasonably fitted with an hyperbolic cosine function,
the asymptotes of these fits do not match: the cell shape is close to a ``minimal'' surface with a different weight on
its curvatures, $\kappa \sig + \kappa_{\perp} \sigO = 0$, where $\kappa_{\perp}$ is the curvature in a plane orthogonal to
the side view. In light of Laplace law, these weights can be interpreted as tensions of different magnitude in the
longitudinal ($\es$ in \fig{fig.atef.sketch}) and orthoradial ($\ep=\en\times\es$) directions. The reason for these different tensions 
and details of this Laplace law are given in the next section. We cannot fit the cell shape with an analytical function matching
this law, firstly because functions solving the corresponding differential equation have never been investigated and do not have
the same properties as the hyperbolic cosine, secondly because the tensions $\sig$ and $\sigO$ actually vary in some measure
along the vertical direction.

From this we can calculate the volume of the cell through time. Here again, the physicist is surprised
to find that the volume defined by \eq{eq.arccircle} and the plate positions $z = \pm h$ is not a constant,
\sfig{fig.evolution}. However, this volume is not the volume of the cell itself but the volume enclosed by
the lateral cell boundaries: indeed, it is observed in side views of cells presenting such a large enclosed 
volume increase that the cell detaches from the microplates in the central region of the contact area 
(\sfig{fig.evolution}), forming a ``pocket'' between the cell membrane and the microplate,
which has every reason to be filled
by culture medium seeping between the adhesions seen in \sfig{fig.shape.TIRF}. Although it was
not possible to track the volume of these pockets through time and compare it to the
enclosed volume variations, we can estimate the energetic cost of the corresponding
water flow: adhesions are
more than 1 $\mu$m apart over 10 to 20 $\mu$m length between the periphery and the
central region where medium pocket is being formed.  Assuming a low
estimate of $N>30$ passages of height 0.1 $\mu$m through which medium can
flow from the periphery to the cell interior, we obtain that the
pressure needed to drive the flow noted in \sfig{fig.evolution} is about $10$ Pa, and
that the power needed for this is of the order $10^{-17}$ W : that is, 2
orders of magnitude smaller than the actomyosin power transmitted to the
microplate and measured in the experiments. The formation of such pockets is thus 
very plausible, and indeed is observed in most experiments when the force is large.

The data we present does not allow to check whether the totality of the change of apparent
volume is due to this water seepage. Therefore, there may be also some volume variations due
to a regulation of cell volume \cite{Jiang-Sun.2013.1} superimposed to the one due to the
formation of the pocket. While the accurate measurement of these variations 
would be important to understand the spreading dynamics of the cells in our setup, the model
below does not require to assume volume conservation in order to predict the vertical deflection
dynamics of the microplates.

\subsubsection{State of stress of the actin cortex in experiments}

Experiments of single cell stretching \cite{Mitrossilis+Asnacios.2009.1} allow to track
simultaneously the geometry and force generated by cells between two microplates
with an arbitrary stiffness, see \fig{fig.atef}, \sfig{fig.shape} and \ref{fig.evolution}. 
In order to check whether the
rheological model developed above can explain the observed cell behaviour, we need
first to calculate the state of stress within the actin cortex from the experimental
observables.

Since it was shown in \citetext{Mitrossilis+Asnacios.2009.1} that the force generation in these
experiments is due to acto\-myosin contraction, we model the cells as an acto\-myosin
surface (shell) surrounding an incompressible but passive cell body (cytosol, nucleus and non-cortical
cytoskeleton), whose mechanical action is solely represented by a homogeneous
internal pressure difference with the medium outside,
$P = P_{\mathrm{cell}} - P_{\mathrm{medium}}$.
Actomyosin being considered here as a thin structure, we perform here the calculations
in terms of a surface tension $w \ts{\sigma}$, where $w$ is the thickness of the acto\-myosin
cortex. $\ts{\sigma}$ is assumed to be a tensor tangential to the cortex and to have no 
variation across $w$.

Because inertia is irrelevant at this scale, the spring force $F = 2k(h_0-h)$
of the microplate device needs to be balanced at any instant by the combination
of the cell body pressure force and the tension force in the cell cortex,
\begin{align}
F = 2\pi R_p\, w\sig\at{z=h} - \pi R_p^2\, P.
\label{eq.force_pressure}
\end{align}
Here, $R_p$ is the radius of the cell at a microplate, and
$\sig$ is the tension of the acto\-myosin cortex along the vertical direction.
It is dependent on $z$, and corresponds to the component $\sig$ along $\es\es$
of the stress tensor of the acto\-myosin tensor, \fig{fig.atef.sketch}. Because of the symmetry and of the
assumption of a thin acto\-myosin cortex, this tensor can only have one other nonzero
component, along the orthoradial direction, $\sigO \ep\ep$.
Using curvilinear coordinates, we can show that 
the force balance in the $\es$ and $\ep$ directions at any height $z$ writes,
\begin{align}
0 = P \en +\div\tsigma 
=
\left(\begin{array}{c}
 P - \kappa w\sig - \frac{\sin \theta}{r} w\sigO
 \\
 w\ded{\sig}{s} + \frac{\cos \theta}{r}w\left(\sig-\sigO\right)
\end{array}\right)
\label{eq.equilirium}
\end{align}
where the first line is Laplace law written with different tensions in the
vertical and orthoradial directions, and the second describes the equilibrium
along direction $\es$.

In order to solve these equations, we need to specify the geometry of the acto\-myosin
walls using \eq{eq.arccircle}. One can then use power series, and
get $\sig=\sig^0 + \sig^1 z^2$ and $\sigO=\sigO^0
+ \sigO^1 z^2$ in a closed form depending on geometrical parameters ($h$, $R_c$,
$\kappa$) and on the pressure difference $P$:
\begin{subequations}
\label{eq.sigs}
\begin{align}
\sig &= \;\sig^0\; + z^2 \frac{\kappa}{2} \left( \sig^0 
        \left(\frac{1}{R_c}+{\kappa}\right) - \frac{P}{w} \right)
        \label{eq.sig}
\\
\sigO&= R_c(P/w-\kappa\sig^0) + z^2 \kappa P/w \left( \kappa {R_c} - \frac{1}{2} \right)
        \label{eq.sigO}
\\
\intertext{with} 
w\sig^0 &= \frac{1}{2} R_c P + \frac{F}{2\pi R_c} \nonumber
\end{align}
\end{subequations}
The presence of $P$ in these equations means that we cannot read directly the
state of stress of the actin cortex from its geometry and the measurement of $F$.
However, if we have e.g.\ an indication on the orthoradial stress $\sigO$, which is
possible when the shape is stationary and thus no dissipation takes place in the 
orthoradial direction, then
both $P$ and $\sigma$ can be determined.
Note that in this section we have not made
use of any assumption on the rheology of acto\-myosin, in particular we have
not used the constitutive \eq{eq.cst} yet: the experimental observations fitted by
a geometry are sufficient to describe the state of stress in the cortex.

\subsubsection{Equilibrium height and force}
\label{sup.three_d.equil} 


We apply the rheological model \eq{eq.cst} in order to predict the
rate of strain $\lded{v_s}{s}$ in the actin cortex along $\es$, and obtain its
value as a power series in $z$ and in function of $F_p$, the geometry and
the model parameters $\Tc$, $\Ta$ and $\Sig$. For this, we need to write the
tensorial constitutive \eq{eq.cst} in curvilinear coordinates, assuming that
$\ts{A}=\es\es + \lambda\ep\ep$, i.e. that the contractile stress in the orthoradial
direction $\ep$ is a fraction $\lambda$ of the contractile stress orthogonal to the
plates:
\begin{subequations}
\begin{align}
\Ta\left(\dedt{\sig} + v_s\ded{\sig}{s} -2\left(\ded{v_s}{s}+\kappa v_n\right)\sig \right) 
        + \sig &
	\nonumber\\
	- 2\Ta E \left(\ded{v_s}{s}+\kappa v_n\right) 
        &= \Sig,
        \label{eq.cst.sig}
        \\
\Ta\left(\dedt{\sigO} - \frac{2}{r} \left({v_s}\cos\theta+v_n\sin\theta\right)\sigO \right) 
        + \sigO&
	\nonumber\\
	- \frac{2\Ta E}{r} \left({v_s}\cos\theta+v_n\sin\theta\right) 
	&= \lambda\Sig.
        \label{eq.cst.sigO}
\end{align}
\end{subequations}

Let us consider a cell that has reached an equilibrium shape, such as the cell
in \sfig{fig.evolution} at time $t=2000$ s. The force plateaus at a value $F_p$, 
the curvature $\kappa$ has reached a
stable negative value, and the equator length $2 \pi R_c$
is steady. Thus $\dot{h}=0$, $\dot{R}_c=0$, $v_n=0$, and
the force does not evolve either ($\dot{F}=0$ and hence $\dedt{\sigma}=0$),
these constitutive  equations simplify to:
\begin{align*}
\Ta v_s\ded{\sigma}{s} + \left(1 -2 \Ta\ded{v_s}{s}\right) \sig - \Tc\Sig \ded{v_s}{s} &= \Sig
\\
\sigO &= \lambda\Sig
\end{align*}
Using the force balance, \eq{eq.sigs}, we can calculate $v_s(z)$ as a function
of $F_p$ :
\begin{align}
v_s =& z\left(-\frac{\phi}{\Tc} + \frac{1-\phi}{2\Ta}\right)\!\left(1 
	+\frac{\kappa z^2}{6 R_c}\!\left(\kappa R_c + \frac{2\Ta+\Tc}{\Tc}\phi\right)\!
	\right)
\label{eq.deps}
\end{align}
where 
$$
\phi = \frac{2 - \Rc \kappa}{2 -\Rc \kappa + 2\frac{\Ta}{\Tc}\left(\lambda + \frac{F_p}{\pi R_c \Sigma}\right)} \,\in\, (0,1]
$$
This flow is the
balance between a contractile term proportional to
$1/\Tc$ and an extensional term in $1/(2\Ta)$. In practice, it is always negative: it
corresponds to a \emph{retrograde flow} that vanishes at the
cell's equator for obvious symmetry reasons, and increases in magnitude with $z$.
It is modulated by geometric factors, but also by the force $F_p$. This retrograde flow is present 
for all values of the external stiffness. 

In order to reach an equilibrium we need the retrograde flow to compensate exactly the addition of new cortex 
through polymerisation at
$z=h$, which means that 
\begin{align}
v_s(h_p)=-\vp.
\label{eq.polym_plateau}
\end{align}
Thus we have a relation for $h_p$ when the geometry is known in terms of $R_c$
and $\kappa$. When the force $F_p$ is low (in the case of vanishing $k$), there is an asymptote value $h_e$ for $h_p$ provided
that it is much smaller than $R_c$, which is found to be the case.
Experiments provide a redundant reading of $h_e$, since the force has to be $2k(h_0 - h_e)$ at low $k$ values. Thus
we also have,
$$
\lim_{k\rightarrow 0} \ded{F_p}{k} = 2(h_0 - h_e).
$$
These consistently give $h_e/h_0 = 0.46 \pm 0.06$. In \eq{eq.deps}, $h_e$ is proportional to $\Tc\vp$ as in the 1D
model, but is modulated both by the curvature (which is observed) and the contractility in the orthoradial direction,
which cannot be accessed.
Using the values $\Tc=521 \pm 57$ s and $\Ta=1186\pm 258$ s obtained from the
dynamics of the 1D model (see \ref{sup.one_d.dyn}), it is found that the model
can predict the cell behaviour only if the orthoradial contractility
$\SigO=\lambda\Sigma$ is significantly lower than $\Sigma$, $\lambda\lesssim
0.5$---else the pressure build-up in the cytosol prevents the cell from
contracting.  We thus take $\lambda=0.5$ and $\vp = 4$ nm/s
which is close to the value found in 1D ($6.5 \pm 1.5$) and in the literature
($4.3 \pm 1.2$ nm/s,  \citetext{Rossier+Sheetz.2010.1}).

There remains one last free parameter in the 3D model, the (surfacic) contractility $w\Sigma$. This can 
be assessed in the limit of infinite microplate stiffness $k$, we find $w\Sigma = 15$ nN/$\micron$. 
Using these values, the 3D model yields a plateau force very close to the one of
the 1D model, see \fig{fig.F_vs_k}.

\ifnum\usebibtex=0
\begingroup
    \fontsize{8pt}{10pt}\selectfont

\endgroup
\else
\fi

\end{supplements}

\ifnum\showlegends=1
\section*{List of SI figures}
\begin{itemize}
\item[S1] (\textit {\subref {fig.shape.circles}}) Light transmission image of a cell spreading on microplates with infinite stiffness $k$ seen from the side. The sequence of shapes assumed by the cell walls in the course of an experiment can be described by arcs of circles. (\textit {\subref {fig.shape.TIRF}}) TIRF visualisation of fluorescent paxillin in a cell spreading on microplates with infinite stiffness $k$ seen from the bottom. The spreading is isotropic, which supports the axial symmetry hypothesis. Adhesion zones are clearly apart from one another, along a circular region of interest we find $N\simeq 40$ adhesion zones separated by paxillin-free passages of average width $1.5\,\micron$ (see \refsup{sup.geometry}). 
\item[S2] Time evolution of the force and geometry of a single cell spreading between microplates of intermediate stiffness $k=176$ nN/$\mathrm {\upmu m}$. Top, the force grows until it reaches a maximum value. Center, concurrently with the force increase, the cell spreads on the microplates, $R_p$ increases. The radius at the equator $R_c$ also increases after a transient decrease. Both stabilise when the force is maximal. As the cell deflects the microplates, its half-height $h$ decreases, however this decrease does not compensate the spreading in terms of (apparent) volume, and the volume $V$ enclosed by the lateral cell surfaces increases more than two-fold. Bottom, transmission images show that this apparent volume increase happens concurrently with the formation of `pockets' (arrow heads) away from the peripheral cell adhesions (\fig {fig.shape.TIRF}) where the cell locally detaches from the microplate. See \refsup{sup.geometry} for details.
\item[S3] Microtubules have a negligible influence on stiffness-dependent force generation. Blue boxes, plateau force exerted by cells in microplate experiments in presence of 1 $\upmu$M Colchicine. Red crosses, control. See \refsup{sup.microtubules} for a discussion.
\item[S4] Plateau force measured for two different cell types, Ref52 fibroblasts and C2-7 myoblasts. (\textit{A}) Plateau force divided by external stiffness, $F/k$, in $nN/(nN/\micron)$, for $k < k_c$. Welch two-sample $t$-test cannot discriminate them, $p$-value $0.083 > 0.05$. (\textit{B}) Plateau force for $k > k_c$, in $nN$ (includes both experiments with side- and bottom-view for Ref52 cells). Welch two-sample $t$-test cannot discriminate them, $p$-value $0.17 > 0.05$. 
\end{itemize}
\fi

\end{multicols}

\setcounter{figure}{0}
\renewcommand{\figurename}{Supplementary Fig.}
\renewcommand{\thefigure}{S\arabic{figure}}
\newlength{\movefigright}
\setlength{\movefigright}{0pt}

\setlength{\figwidth}{.8\textwidth}
\begin{figure}[b]
\centering
\begin{tabular}{ll}
\begin{subfigure}[t]{.5\figwidth}
\hspace*{\movefigright}
\includegraphics[height=.75\textheight]{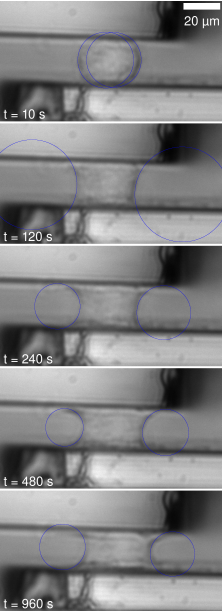}
\caption{}
\label{fig.shape.circles}
\end{subfigure}
\begin{subfigure}[t]{.25\figwidth}
\includegraphics[height=.75\textheight]{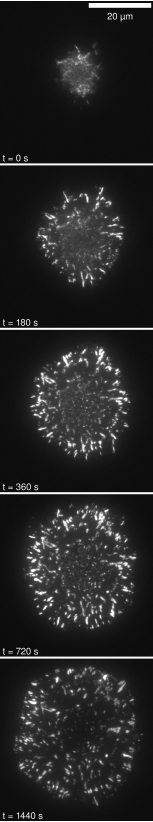}
\hspace*{-.4\figwidth}
\caption{}
\label{fig.shape.TIRF}
\end{subfigure}
\end{tabular}
\caption[
(\textbf{\subref{fig.shape.circles}}) Light transmission image of a cell spreading
on microplates with infinite stiffness $k$ seen from the side.
The sequence of shapes assumed by the cell walls in the course of an experiment can
be described by arcs of circles.
(\textbf{\subref{fig.shape.TIRF}}) TIRF visualisation of fluorescent paxillin in a cell spreading
on microplates with infinite stiffness $k$ seen  from the bottom.]{}
\label{fig.shape}
\end{figure}

\begin{figure}[t]
\centering
\includegraphics[width=.7\textwidth]{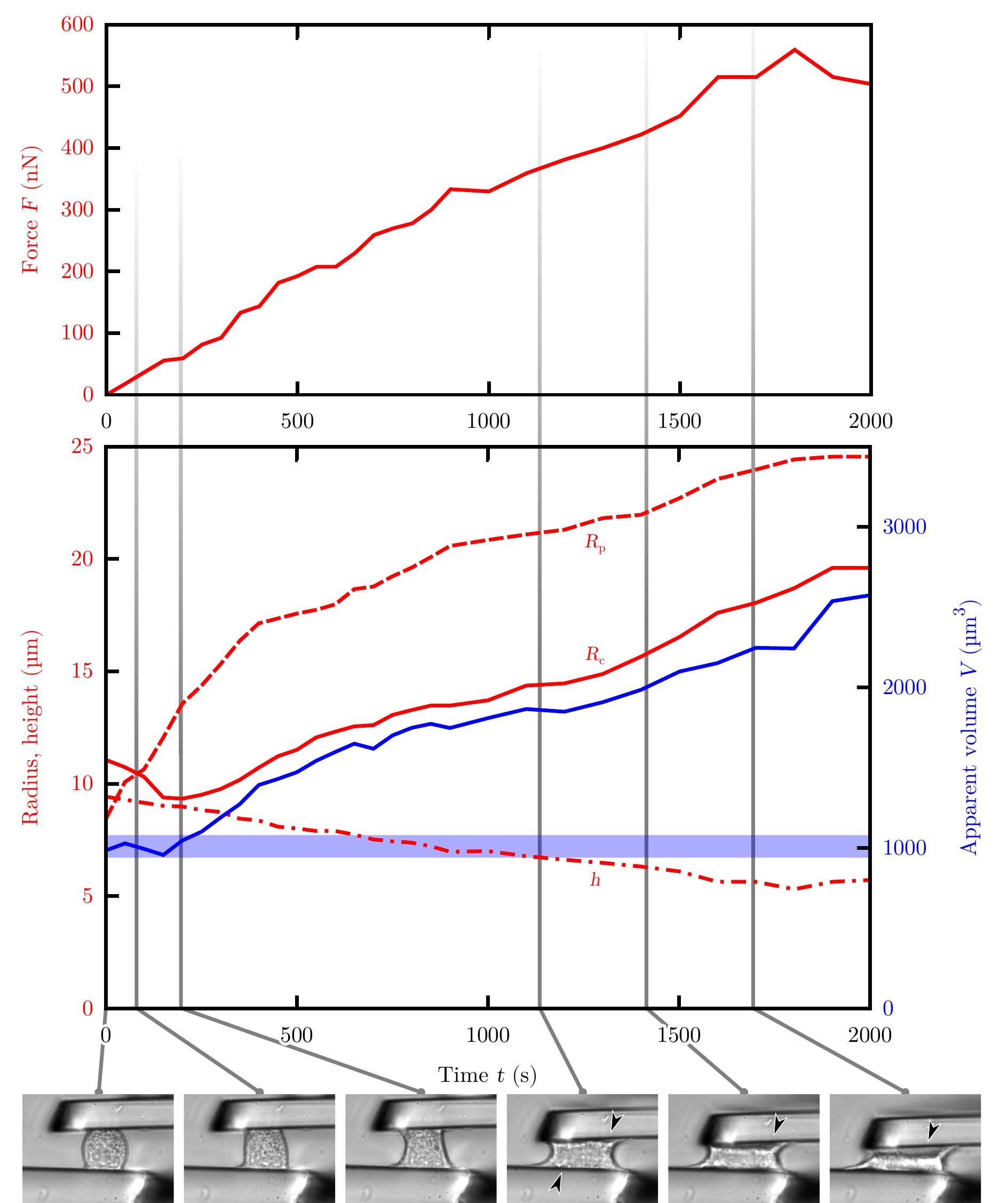}
\caption[
Time evolution of the force and geometry of a single cell spreading between microplates
of intermediate stiffness $k=176$ nN/$\micron$. Top, the force grows until it reaches a
maximum value. Center, concurrently with the force increase, the cell spreads on the microplates,
$R_p$ increases. The radius at the equator $R_c$ also increases after a transient decrease.
Both stabilise when the force is maximal. As the cell deflects the microplates, its half-height $h$
decreases, however this decrease does not compensate the spreading in terms of volume, and the
volume $V$ enclosed by the lateral cell surfaces increases more than two-fold. Bottom, transmission
images show that this apparent volume increase happens concurrently with the formation of `pockets' (arrow heads)
away from the peripheral cell adhesions (\fig{fig.shape.TIRF}) where the cell locally detaches from
the microplate.]{}
\label{fig.evolution}
\end{figure}

\begin{figure}[tp] 
\begin{center} 
\centering
\includegraphics{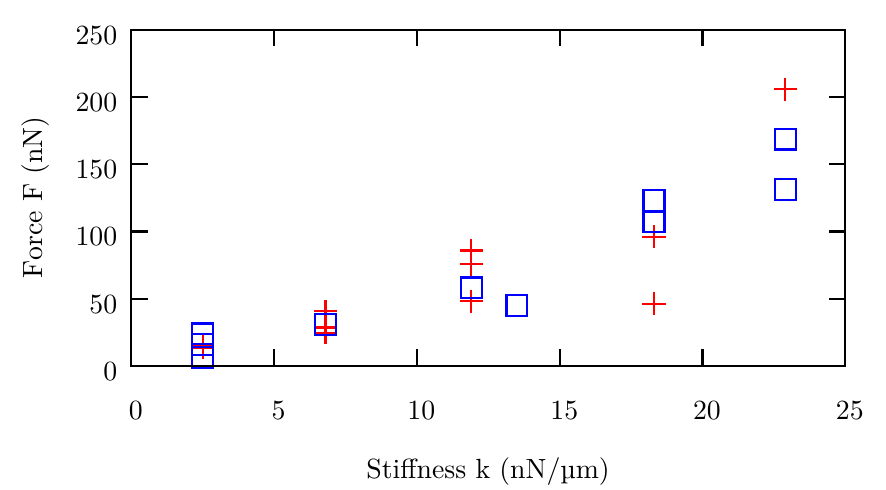}\\ 
\caption[]{}
\label{fig.microtubules}
\end{center}
\end{figure}

\begin{figure}[tp]
\setlength{\unitlength}{\linewidth}
\centering%
\begin{minipage}[t]{.4\linewidth}
\leavevmode\put(.11,.4){\leavevmode\makebox(0,0){$F_p/k$}}
\includegraphics[width=\linewidth]{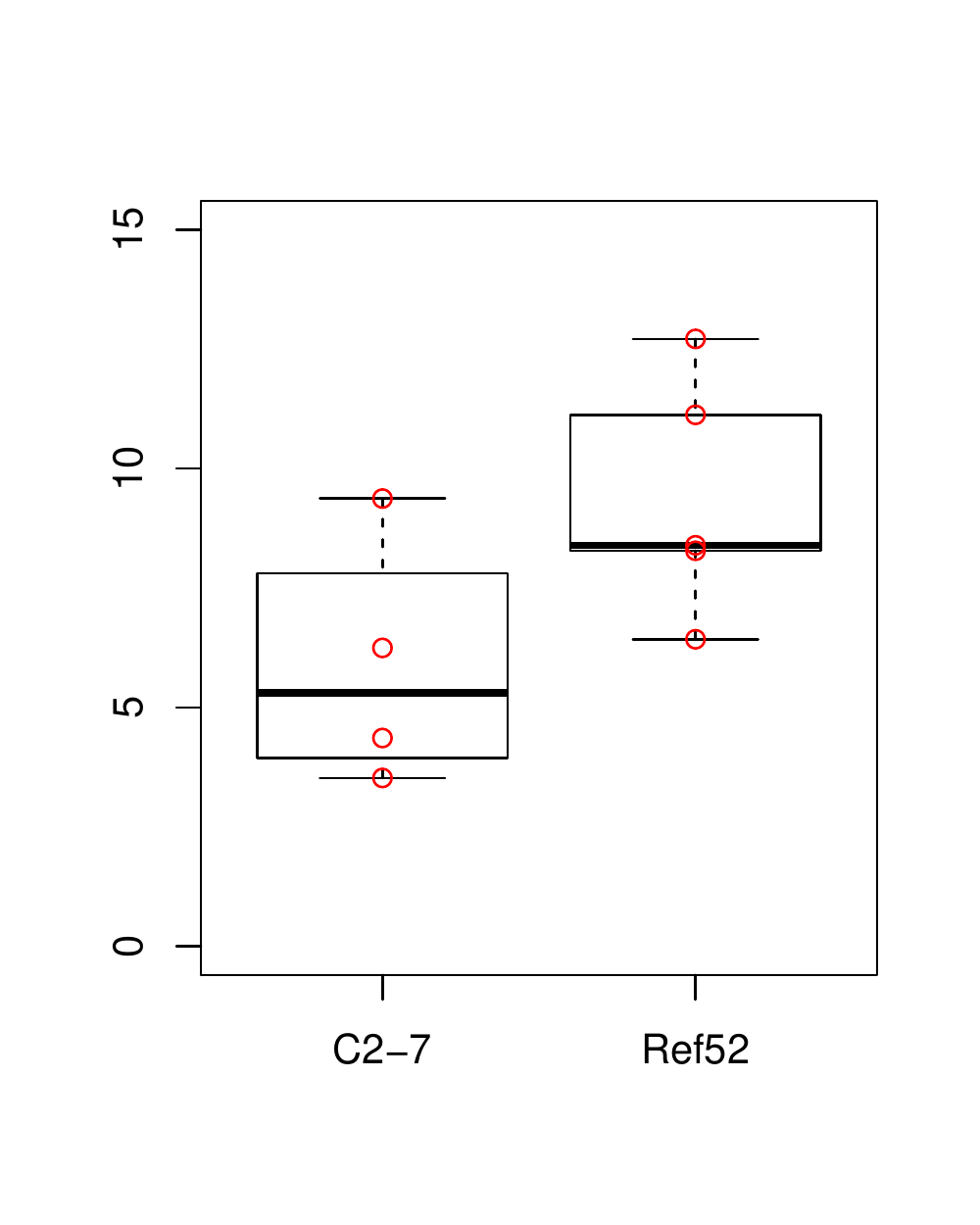}
\\\centering%
\textit{a}
\end{minipage}
\begin{minipage}[t]{.4\linewidth}
\leavevmode\put(.11,.4){\leavevmode\makebox(0,0){$F_p$}}
\includegraphics[width=\linewidth]{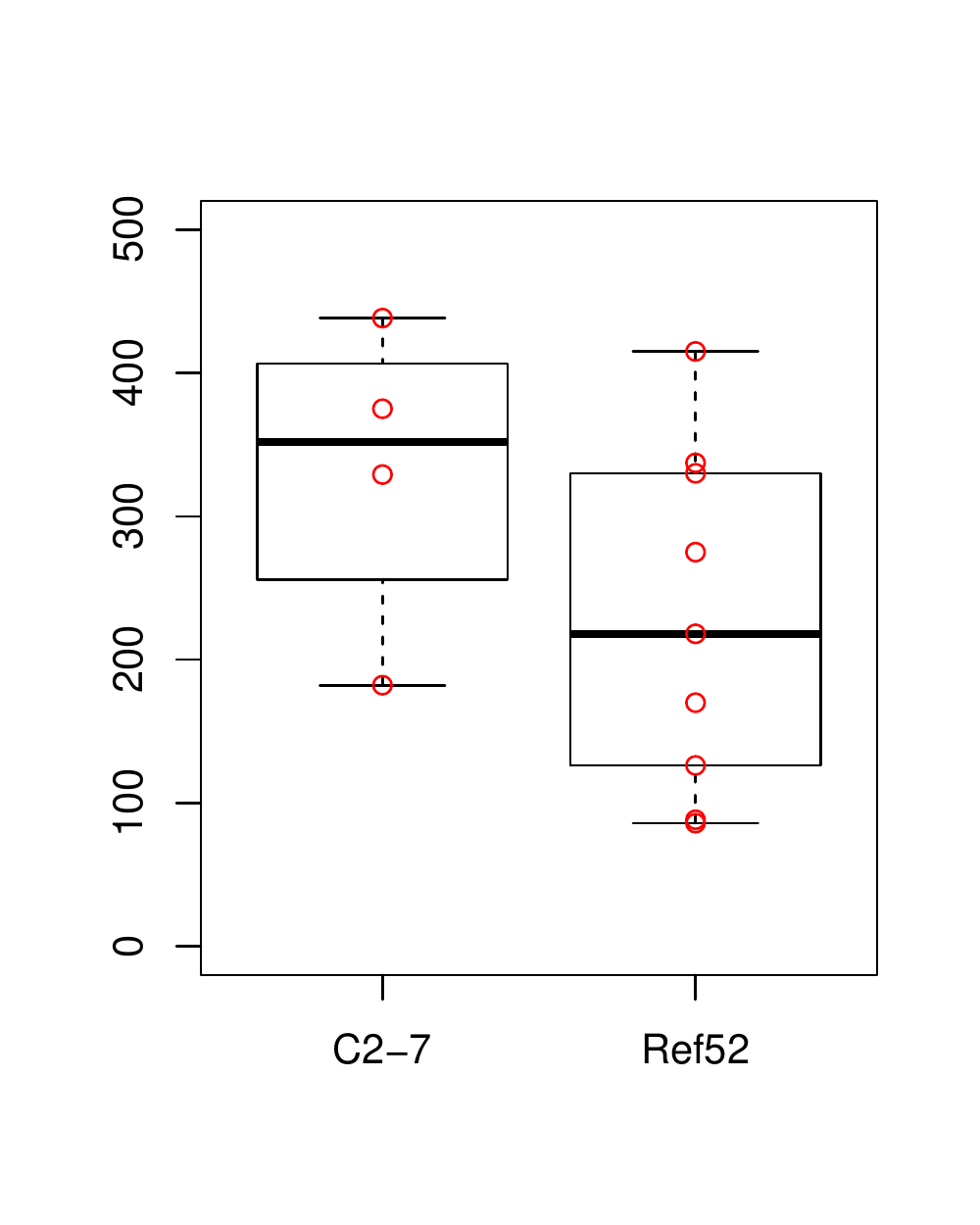}
\\\centering\textit{b}
\end{minipage}
\centering%
\caption{}
\label{fig.compare_cell_types}
\end{figure}

\end{document}